\documentclass[twocolumn]{aa}
\usepackage{graphicx,psfig}
\begin{document}

  \title{Chemical compositions and plasma parameters of planetary nebulae 
          with Wolf-Rayet and {\it wels} type central stars
    \thanks{Based on observations obtained at the European Southern
        Observatory (ESO), La Silla, Chile }}
    \author{P.Girard \inst{1,2}
       \and J.K\"oppen \inst{2,3,4} 
       \and A.Acker \inst{2}
  }
  \offprints{girard@obs.u-bordeaux1.fr}      
       
  \institute{Observatoire Aquitain des Sciences de l'Univers, L3AB, 
       2 rue de l'Observatoire, BP 89, 33270 Floirac, France
  \and	
      Observatoire Astronomique de Strasbourg,
     11 rue de l'Universit\'e, 67000 Strasbourg, France
  \and 
     Institut f\"ur Theoretische Physik und Astrophysik,
     Universit\"at Kiel, D-24098 Kiel, Germany
  \and International Space University, Parc d'Innovation,
     F-67400 Illkirch, France}

  \titlerunning{Planetary nebulae with [WR] central stars}
  \date{Received / accepted}
%
%
%
   \abstract
  {} 
 {Chemical compositions and other properties of planetary nebulae 
    around central stars of spectral types [WC], [WO], and $wels$ are compared
    with those of `normal' central stars, in order to clarify the evolutionary status 
    of each type and their interrelation.}
 {We use plasma diagnostics to derive from optical spectra the plasma parameters
  and chemical compositions of 48 planetary nebulae. We also reanalyze the published
  spectra of a sample of 167 non-WR PN. The results as well as the 
  observational data are compared in detail with those from other studies of the 
  objects in common.}
 {We confirm that [WC], [WO] and $wels$ nebulae are very similar to those 
   `normal' PN: the relation between [N~II] and [O~III] electron temperatures, 
   abundances of He, N, O, Ne, S and Ar, and the number of ionizing photons 
   show no significant differences. 
   However, some differences are observed in their infrared (IRAS) properties. 
   $wels$ nebulae  appear bluer than [WR] PN.
   The central star's spectral type is clearly correlated with electron density, 
   temperature and excitation class of the nebula, [WC] nebulae tend to be smaller 
   than the other types. All this corroborates the view of an evolutionary sequence 
   from cool [WC~11] central stars inside dense, low excitation nebulae towards 
   hot [WO~1] stars with low density, high excitation nebulae. The $wels$ PN, however, 
   appear to be a separate class of objects, not linked to WRPN by evolution: nebular 
   excitation, electron temperature and density, and the number of ionizing photons all 
   cover the whole range found in the other types. Their lower mean N/O ratio and slightly 
   lower He/H suggest progenitor stars less massive than for the other PN types. 
   Furthermore, the differences between results of different works are dominated by
   the differences in observational data rather than differences in the analysis methods} 
  {} 
  \keywords{planetary nebulae: abundances 
 	    -- stars: abundances
 	    -- stars: evolution 
 	    -- stars: Wolf-Rayet
 	    -- ISM: planetary nebulae 
 	    -- stars: AGB}

  \maketitle  

  \section{Introduction}

   Planetary nebulae (PN) are the highly visible transitionary phase in the life of 
   intermediate mass stars on their evolution from the asymptotic giant branch 
   to their final destination, the white dwarfs. Among the 1300 objects known in
   our Milky Way (Acker et al. 1992, 1996b), there are about 6\% whose central stars 
   show broad emission lines, characteristic of the [WR] spectral type, and most likely 
   produced by a massive continuous mass loss from the central star. Whether these 
   objects form a group or an evolutionary phase or evolutionary sequence distinctly 
   different from the other, `normal' PN, is still not fully understood.

   That PN with [WR] central stars do not seem to have properties which differ very
   much from normal, non-WR objects was shown by G\'orny \& Stasi\'nska (1995) who found
   that bipolar nebulae constitute about 20 \% of the total in both WR and non-WR
   objects. Also, the distribution of He/H, N/O and C/O abundance ratios are the same 
   in either group. Both aspects indicate that the WR phenomenon does not 
   preferentially occur in more massive central stars, hence more massive progenitor
   stars. Acker et al. (1996a), G\'orny \& Tylenda (2000) and Pe\~na et al. (2001) 
   showed that the majority of [WR] PN seems to form an evolutionary sequence from 
   late-type [WC] inside high-density nebulae to early-type [WO] with low-density
   nebulae.
  
   In their recent quantitative classification of central stars, Acker \& Neiner (2003) 
   distinguish two sequences from the late-type [WC] to the early-type [WO]:
   The spectra of hot [WO~1...4] types are dominated by the highly ionised oxygen 
   lines, while those of the cooler [WC~4...11] types are marked by carbon lines. 
   There seems to be an evolutionary sequence from [WC~11] to [WO~1]. The {\it wels} 
   (weak emission line stars) objects differ from other central stars 
   (Tylenda, Acker \& Stenholm 1993) and are not part in this classification scheme.

   More recently, G\'orny et al. (2004) found that the proportion of WRPN in the 
   Galactic bulge is about 15\%, significantly larger than in the disk (about 6\%).
   Among the bulge WRPN about 47\% are of type later than [WC~9]. In the disk
   this fraction amounts to only 17\%. However, this finding is strongly sensitive
   to observational selection effects. They confirm the strong trend for the density 
   to decrease towards early-type [WC]. Oxygen abundances in bulge WRPN are 
   found to be the same as in the bulge non-WR, and  the N/O distributions of PN 
   in disk, bulge and of [WR] type are similar.

   Another age indicator is the dust temperature derived from near and mid
   infrared data, indicating an evolutionary sequence from carbon AGB stars
   to [WC] PNe (Acker et al. 1996a). G\`orny et al. (2001) found that
   a sizeable fraction of WRPN seems to contain hot dust (1000-2000 K),
   probably in the winds of the central stars. The mean dust temperature decreases 
   towards late-type [WC], in line with an evolutionary sequence 
   from [WC11] to [WC2] ([WO2]).
     
   The present paper presents the study of a homogeneous sample
   of nebulae around nuclei with emission lines, using the high S/N spectra 
   which had been used to study the [WR] central stars by Acker \& Neiner (2003).
   These objects constitute a sample similar to that of Pe\~na et al. (2001), 
   and it seems interesting to confirm (or not) the conclusions of these authors 
   as well as those  by G\'orny et al. (2004).
  
  After presenting the observational material in section 2 and describing the
  analysis of the spectra in section 3, we study the chemical compositions
  of the various types of nebulae in section 4. We also compare the different 
  groups with respect to other global relations in section 5, investigate the
  relation between central stars and the nebulae (section 6), and analyse the infrared
  properties (section 7). In appendices \ref{s:comparemethods} and \ref{s:comparedata}
  we study the influences on the abundances due to differences in observational 
  data and analysis methods.

\section{Observational material}

  90 spectra of PN around central stars with emission lines were obtained in 
  March 1994, July 1994, and July 1995 with the Boller \& Chivens spectrograph 
  and the CCD detector on the 1.52m telescope at ESO, Chile. The wavelength 
  range is from 3700~\AA\ to 7500~\AA , with a spectral resolution of 1500 (more 
  details can be found in Acker \& Neiner 2003). Exposure times of 40 min and 5 min 
  were employed, giving signal to noise ratios of better than 30 for most objects. 
  For the analyses, the 48 objects with the best spectra were selected.
 
\begin{figure}[!ht]
      \includegraphics[trim = 70 70 70 70, scale=0.35, angle=270]{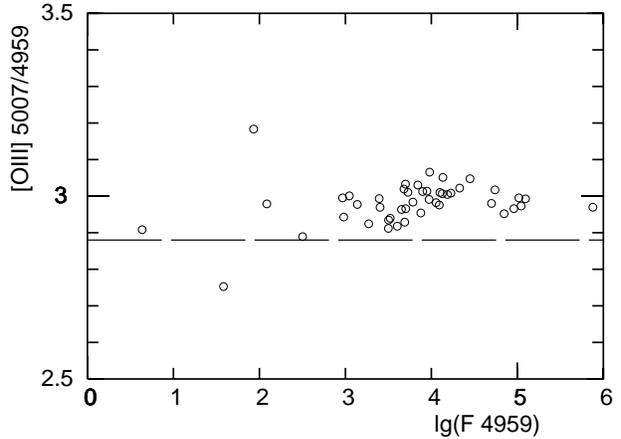} 
       \caption[]{The intensity ratio of the [O~III] 5007 and 4959 \AA\ lines as
                  a function of the flux (arbitrary units) in the 4959 \AA\ line. 
                  The horizontal line indicates the value expected from the
                  atomic data compiled by Mendoza (1983)}
     \label{f:oiiiratio}   
   \end{figure}

  The measured line fluxes and dereddened intensities from the spectra of 48 planetary 
  nebulae with emission line-stars used in this work are available in 
  electronic form at the CDS\footnote{http://cdsweb.u-strasbg.fr/}.
  The quality of the data and their reduction can be assessed by looking at the 
  line ratio of the [O~III] 5007 and 4959~\AA\ doublet which is independent of 
  physical conditions in the nebula. Figure \ref{f:oiiiratio} shows that in all 
  objects values very close to 3.0 are found, but slightly above the theoretical
  value from Mendoza (1983), as expected (cf. Acker et al. 1989). 
  Only a few objects with faint spectra exhibit somewhat larger deviations.

\begin{figure}[!ht]
     \includegraphics[trim = 70 70 70 70, scale=0.35, angle=270]{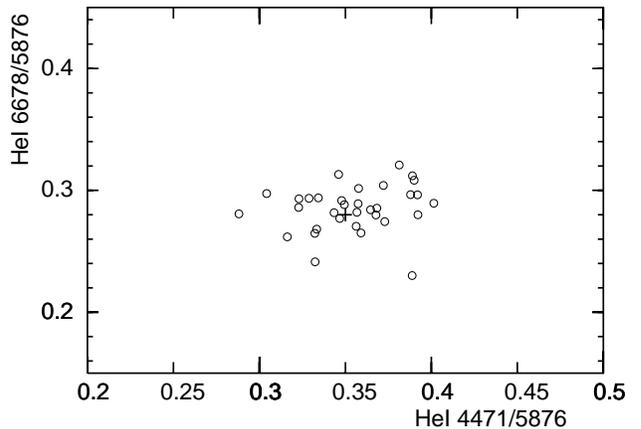} 
      \caption[]{The intensity ratios of the HeI 5876, 4471, and 6678~\AA\ lines.
                 The plus sign marks the low density value for $T_e = 10^4$~K.}
    \label{f:helines}   
  \end{figure}

  Another check is provided by the intensity ratios of the de-reddened He~I lines 
  at 5876, 4471, and 6678~\AA\ which are not very sensitive to nebular conditions. 
  All 33 objects which contain the three lines are found within 20 percent of the 
  theoretical values (Fig. \ref{f:helines}).
   
\begin{figure}[!ht]
     \includegraphics[trim = 70 70 70 70, scale=0.35, angle=270]{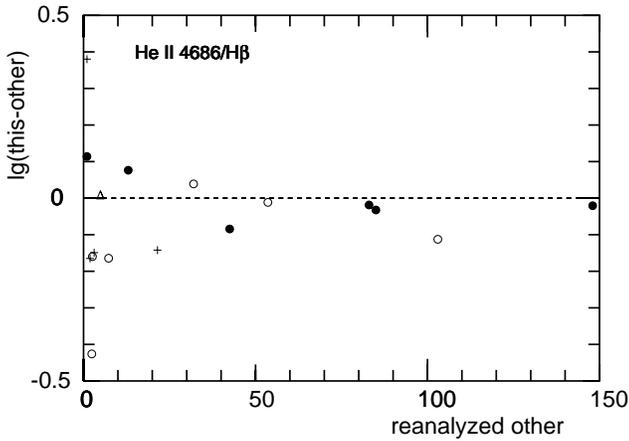} 
      \caption[]{The relative difference in intensity of the He~II 4686\AA\ 
                 line of our measurements compared to those of other works: 
                 Aller \& Keyes (1987, AK87, $+$),
                 Kingsburgh \& Barlow (1994, KB94, $\circ$),
                 Cuisinier, Acker \& K\"oppen (1996, CAK96, $\bigtriangleup$),  
                 and Pe\~na, Stasi\'nska \& Medina (2001, PSM01, $\bullet$)}
    \label{f:cpaheii}   
  \end{figure}

  Since a number of the objects have been observed by other authors, 
  we compared our data in detail with previous measurements. 
  The He~II 4686~\AA\ line covers a large range in values among the objects.
  As shown by Fig. \ref{f:cpaheii}, our (dereddened) intensities 
   agree very 
  well with those obtained by other authors. In particular, the agreement 
  with the most recent study of Pe\~na et al. (2001, PSM01) is better than 0.1 dex.  As one should 
  expect, the scatter increases towards fainter intensities.
  
\begin{figure}[!ht]
      \includegraphics[trim = 70 70 70 70, scale=0.35, angle=270]{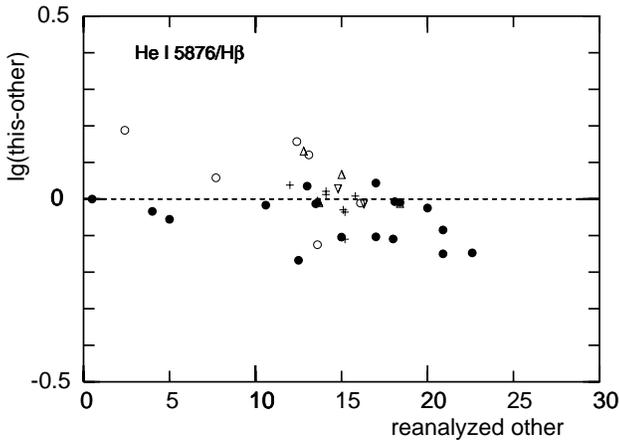} 
       \caption[]{The intensity of HeI 5876~\AA\ line compared to the 
                  value obtained by other works. Symbols are as in
                  Fig. \ref{f:cpaheii} plus Cuisinier et al. (2000, CMAKS00, $\bigtriangledown$)}
     \label{f:cpahei}   
  \end{figure}

  A comparison of the He~I 5876~\AA\ lines (Fig.\ref{f:cpahei}) shows
  a good agreement with other studies. However, one notes a trend, in
  that our intensities tend to be larger than those of Kingsburgh \& Barlow (1994, KB94), but
  smaller than those by PSM01 which is especially obvious 
  for larger intensities.
  Objects with large deviations are the [WO]s PNG~$002.4+05.8$, 
  PNG~$003.1+02.9$, PNG~$017.9-04.8$, PNG~$278.1-05.9$, 
  the [WC]s PNG~$006.8+04.1$, PNG~$027.6+04.2$, PNG~$048.7+01.9$,
  and the {\it wels} PNG~$010.8-01.8$.
  Since [WO] nebulae exhibit no stellar He~I lines (cf. Acker \& 
  Neiner 2003), it is unlikely that our measurements are affected
  by confusion with the stellar spectrum. Furthermore, inspection 
  of the other He~I lines (4471, 6678, 7065, and 5015\AA) shows that 
  these line ratios are in good agreement with the intensity of the 
  5876 line.   

\begin{figure}[!ht]
     \includegraphics[trim = 70 70 70 70, scale=0.35, angle=270]{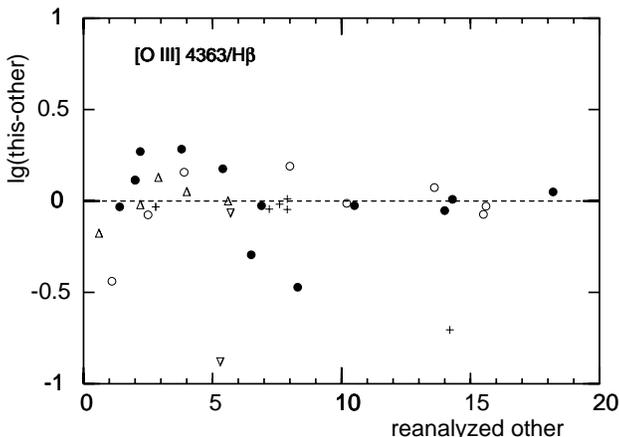} 
       \caption[]{The measured intensity of the [O~III] 4363~\AA\ line compared
                  to the values obtained by the other works. Symbols are as 
                  in Fig. \ref{f:cpahei}}
     \label{f:cpaoiiia}   
  \end{figure}

  The [O~III] 4363~\AA\ diagnostic line is of similar strength as 
  the He~I 5876~\AA\ line. As shown in Fig. \ref{f:cpaoiiia}, our measurements 
  are in good to fair agreement with the intensities obtained by the other
  studies. For intensities below 10 percent of H$\beta$\ the deviations
  with PSM01 are as large as a factor of two, in either direction.
  A similar finding is obtained for the [S~II] 6717/31~\AA\ doublet.

  As a consequence of the difference in He~I line intensities, our helium 
  abundances differ slightly from the results of PSM01. This is addressed
  in more detail in Appendix \ref{s:comparedata}. However, it
  does not affect the findings of this paper with respect to the evolutionary 
  status of the [WR] PN.

\subsection{A sample of 'normal' PNe}

  To compare the sample of [WR] nebulae with `normal' PN, we selected
  the objects from the following works: 
  Aller \& Czyzak (1983, hereafter AC83), 
  Aller \& Keyes (1987, AK87), Cuisinier, Acker \& K\"oppen (1996, CAK96), 
  Cuisinier et al. (2000, CMAKS00) 
  and Kingsburgh \& Barlow (1994, KB94).
  Although the image tube data from   AC83 and AK87 
  might be considered less reliable than present CCD data, such a suspicion 
  is not too evident in individual comparisons. Moreover, these works 
  constitute a large sample of the brighter PN, together with 
  KB94. We include CAK96 who focussed on objects
  high above the Galactic plane as well as the sample of Bulge nebulae
  by CMAKS00. The inclusion of these samples 
  does not constitute a major bias, since in our sample of [WR] objects 
  there also is no selection against nebulae outside the disk.

  To avoid possible influences by the different analysis methods employed by 
  the other authors, we re-analyzed all the other spectra by our method using 
  the same criteria. This includes using only the optical lines in the spectra 
  of KB94. All together, this sample provides 167 objects with spectra of 
  comparable quality which had been obtained for determinations of 
  chemical compositions.  

\subsection{Comparison with other works}

  There are 3 nebulae in common with AC83, 7 with AK87, 8 with KB94,
  6 with CAK96, 2 with CMAKS00, and in particular 17 with PSM01. 
  This offers an opportunity to clarify to what extent the results from 
  different studies are subject to differences in the measured line
  fluxes or to the adopted abundance determination methods. 
  In Appendix \ref{s:comparemethods} we compare the results
  obtained from reanalysis of the data by HOPPLA with the original 
  values. The differences in the observational data are addressed in 
  Appendix \ref{s:comparedata}.
 
\section{Plasma analysis of the nebular spectra}

  Extinction constants, electron temperatures and densities, and
  elemental abundances are determined with the computer programme
  HOPPLA (see Acker et al. 1991, and K\"oppen et al. 1991).
  The spectra are interpreted by the technique of plasma diagnostics, 
  viz. assuming that all lines are produced in an isothermal gas at 
  uniform density and ionization level. In the first step, the reddening 
  correction, derivation of electron temperature and density, as well
  as the optical depth of the He~I 3888~\AA\ line (for self-absorption
  in the He~I lines) are performed. These steps are repeated several 
  times, until the values converge. 

  The excitation class (EC), absolute H$\beta$ fluxes seen through the 
  spectrograph aperture, extinction constants $c$, electron temperatures 
  and densities for all selected 48 PN are compiled in Table \ref{t:plasma}.
  The column `Q' gives an indication of the overall quality of the analysis. 
  `A' means that all diagnostic lines are present, `B' that the density could 
  not be reliably determined, and `C' that the electron temperature could not 
  be derived.

\begin{figure}[ht]
    \includegraphics[trim = 0 0 0 0, scale=0.35, angle=90]{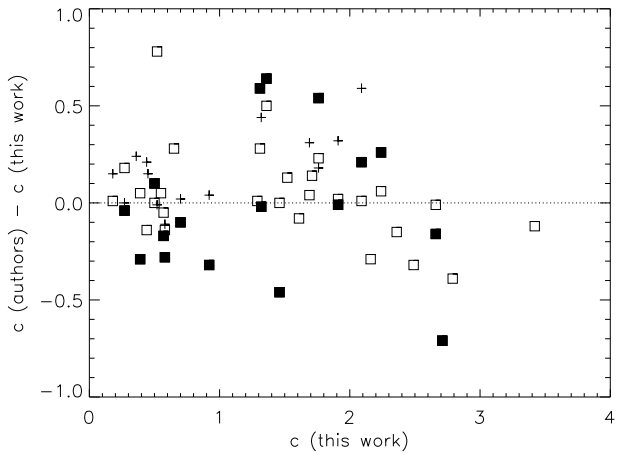} 
      \caption[]{The extinctions found in this work compared to other
                 works for the same nebulae. 
                 Open squares refer to Acker \& Neiner (2003) 
                 (from Crowther et al. 1998 and Tylenda et al. 1992),
                 filled squares to PSM01, and plus-signs to AC83, 
                 Aller \& Keyes (1980, 1987),  Aller et al. (1981, 1986),
                 Barker (1978), KB94,
                 and Shaw \& Kaler (1989)}
   \label{f:exc}  
  \end{figure}

  The extinction measure $c$ is obtained from the decrement of all the 
  observed Balmer lines relative to the one computed from Brocklehurst (1972) 
  for case B. To find the optimum value, the error from each line is weighted 
  with the square of the observed line flux. This weighting had been chosen
  to be able to deal also with rather noisy spectra; for the present data
  this suppresses the noise in the blue region of the spectrum, by giving 
  a strong weight to the H$\alpha$/H$\beta$ ratio. In Fig. \ref{f:exc} we compare 
  our values of the extinctions against the results found by other authors. For 
  most of the objects we obtain good agreement; yet for 
  several nebulae rather discrepant values have been reported, most probably 
  due to the difficulty to separate the nebular from the stellar emission lines. 
  Notable exceptions are collected in Table \ref{t:extinc}.

  \begin{table}\centering
     \caption[]{Nebulae whose extinctions found in this work and in the 
                literature show strong differences. Other works are
                indicated by their abbreviations (e.g. KB94 for
                Kingsburgh \& Barlow 1994) used in the References} 
     \label{t:extinc}
     \begin{tabular}{llll}
     \noalign{\smallskip} \hline \noalign{\smallskip}
      PN G & this work & other works & \\
    \noalign{\smallskip} \hline
     \noalign{\smallskip}
   $002.2-09.4$ &  0.39  & 0.44 TASK92, & 0.1 PSM01\\
   $004.9+04.9$ &  1.46  & 1.46 TASK92, & 1.0 PSM01\\
   $009.4-05.0$ &  0.92  & 0.96 AK87,   & 0.6 PSM01 \\
   $017.9-04.8$ &  0.52  & 0.51 SK89,   & 1.30 TASK92 \\ 
   $061.4-09.5$ &  0.00  & 0.09 AC83,   & 0.27 KB94,\\
                &        & 0.45 CMB98,  & 0.23 PSM01 \\
   $292.4+04.1$ &  0.68  & 0.38 TASK92, & 2.7: KB94 \\
   $300.7-02.0$ &  2.49  & 2.17 TASK92  & \\
   $331.3+16.8$ &  0.36  & 0.64 KB94    & \\
   $358.3-21.6$ &  0.18  & 0.33 AKF86   & \\
     \noalign{\smallskip} \hline
     \end{tabular}
  \end{table}

\begin{figure}[!ht]
      \includegraphics[trim = 70 70 70 70, scale=0.35, angle=270]{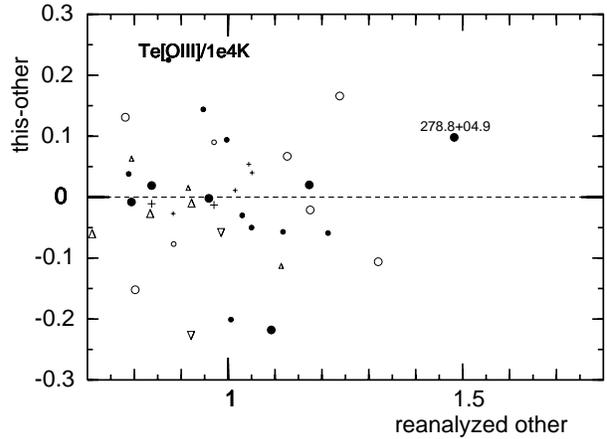} 
       \caption[]{Comparison of the [O~III] electron temperatures
                  derived in this work and determined by other authors
                  for the same objects; symbols are the same as in 
                  Fig.\ref{f:cpaheii}. Smaller symbols indicate that
                  in the other work the values were marked as uncertain}
     \label{f:cpateo}   
   \end{figure}

  Electron temperatures $T_e$ are derived from line intensity ratios 
  of [O~III] (4959~\AA\ + 5007~\AA )/4363~\AA\
  and [N~II] (6548~\AA\ + 6584~\AA )/5755~\AA . 
  We note that the high [N~II] temperature in PNG~$337.4+01.6$ is the 
  consequence of its density to be close to the high density limit 
  of the [S~II] lines. Figure \ref{f:cpateo} shows the comparison of 
  our values of the [O~III] temperature for the objects common with 
  other authors. There is an overall good agreement and no obvious 
  systematic offset. In particular, we confirm the rather high 
  temperature in PNG~$278.8+04.9$ as found by PSM01.

\begin{figure}[ht]
    \includegraphics[trim = 70 70 70 70, scale=0.35, angle=270]{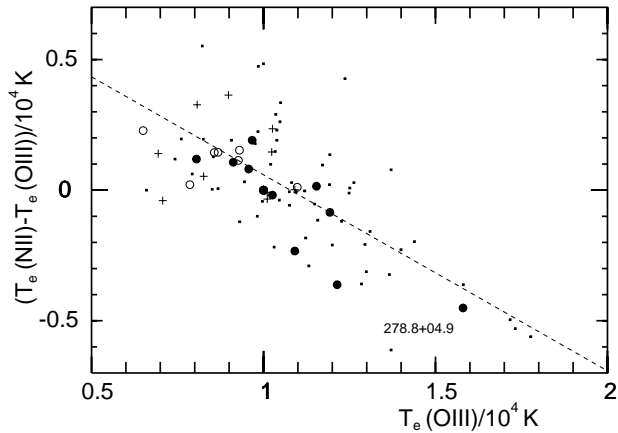} 
    \caption[]{Relation between difference of electron temperatures from [N~II] 
               and [O~III] line ratios with the [O~III] electron temperatures
               for PN with central stars of types [WO] (filled circles),
               [WC] (open circles), {\it wels} (plus-signs), and any other (small dots).
               The dotted line indicates the regression line obtained from the
                `normal' PN only.}
    \label{f:diffte}  
  \end{figure}

  Figure \ref{f:diffte} shows the relation between the  difference between
  [N~II] and [O~III] temperatures with [O~III] temperatures for the 
  [WR] and {\it wels} PN in comparison with the data for the non-WR PN. 
  In nebulae of low electron temperature, the  [N~II] temperature exceeds
  the one in [O~III], but in nebulae of high electron temperature the [N~II]
  temperature is lower. From 74 `normal' PN one obtains a very tight
  correlation, with a coefficient of $-0.74$:
  $$
  {\scriptstyle     (T_e({\rm [N~II]})- T_e({\rm [O~III]})  =  (8090 \pm 923) \\
                     - (0.750 \pm 0.080) \times T_e({\rm [O~III]}) (1)}
  $$
  which is displayed in Fig. \ref{f:diffte}. 
  Taking all [WO] and [WC] PN together, one obtains from 24 objects
  a similarly tight correlation (coefficient $-0.86$):
  $$
  {\scriptstyle   (T_e({\rm [N~II]})- T_e({\rm [O~III]})  =  (7780 \pm 995) 
       - (0.770 \pm 0.97) \times T_e({\rm [O~III]}) (2)}
  $$
  which  is nearly identical. 
  Taking the only 12 [WC], the slope is flatter $-0.509\pm 0.144$ and a 
  correlation coefficient of $-0.74$ is found; from 11 [WO] PN one gets a slope
  of $-0.854 \pm 0.174$ and a coefficient of $-0.85$.
  Thus the [WR] objects follow rather closely the relation seen among the
  `normal' PN. Only a larger sample would permit to judge whether the 
  difference between [WO] and [WC] is really significant.  
  As is already apparent from their scattered positions in Fig. \ref{f:diffte}, 
  the 10 {\it wels} PN do not yield a significant correlation.
  The strong tendency for the `normal' PN objects to have larger [N~II]
  temperatures has already been found by G\'orny et al. (2004) but not for the [WR] objects.
  From their Fig. 2 one also notes a tendency that in the hotter
  nebulae [N~II] temperatures are lower than the ones from [O~III].

\begin{figure}[ht]
    \includegraphics[trim = 70 70 70 70, scale=0.35, angle=270]{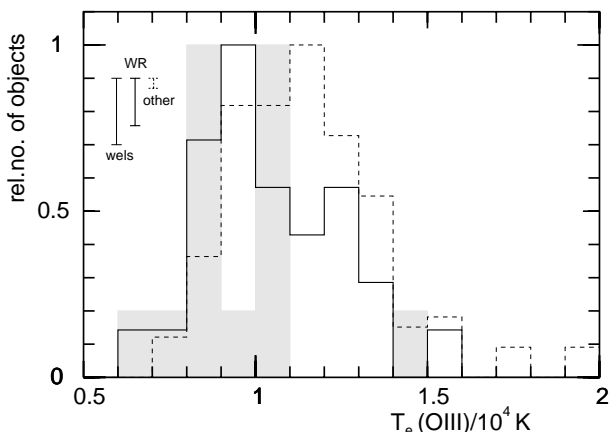} 
    \caption[]{Histograms of the [O III] electron temperature for the
               [WR] objects (full line), the {\it wels} PN (shaded area) and the 
               objects with `normal' central stars
              (dashed). The vertical error bars indicate the contribution
               by a single object.}
    \label{f:tehisto}  
  \end{figure}

  The lower electron temperatures in the [WR] objects are also
  evident in the histogram (Fig. \ref{f:tehisto}). The average temperature
  is $10440 \pm 390$~K from 28 [WR] objects, somewhat lower than 
  $11680 \pm 200$~K from the 164 `normal' PN; the dispersions are
  also quite similar: $2070 \pm 280$~K and $2620 \pm 140$~K, respectively.
  Much more different are the {\it wels} nebulae: with a single exception of 
  PNG~$331.3+16.8$, the 14 objects have [O~III] temperatures in the narrow 
  range between about 8000 and 11000~K, as is also apparent in 
  Fig. \ref{f:diffte}. The mean is $9390 \pm 460$ and the dispersion 
  $1740 \pm 330$~K, which includes the singular PNG~$331.3+16.8$. 

  Inspection of the temperatures and excitation classes shows a mild
  tendency for both [N~II] and [O~III] temperatures to increase with 
  increasing excitation class, but no significant difference could be found 
  in the behaviour of [WR] and normal PN. PNG~$278.8+04.9$ remains 
  exceptional in having one of the highest [O~III] temperatures found 
  among any PN; 
   the other objects seen in Fig. \ref{f:diffte} in its vicinity
  are PNG~$086.5-08.8$ (data from AC83),  PNG~$245.4+01.6$, 
  PNG~$261.6+03.0$, and PNG~$275.8-02.9$ (from KB94).

  Electron densities $n_{e}$ are determined from the intensity ratios of the 
  doublet [S~II] 6717/6731~\AA . If the [S~II] lines cannot be measured, we use 
  the [Cl~III] 5517/5537~\AA\ ratio (indicated by `C' in Table \ref{t:plasma}). 
  In several nebulae the [S~II] line ratio is close to the high-density
  limit. PNe whose density exceeds 20,000 cm$^{-3}$ are marked by a `H'
  in Table \ref{t:plasma} and this adopted value is used in further analysis.

  \begin{table*}[!ht]\centering
   \caption[]{Plasma parameters of the planetary nebulae of this sample. The
                   spectral types are from Acker \& Neiner (2003). Given are: 
		   overall quality of the analysis (see the text),
		   excitation 
                   classes EC following Aller's (1956) system, H$\beta$ fluxes in 
                   erg s$^{-1}$ cm$^{-2}$, extinction constant $c$, electron temperatures
                   from the [O~III] and [N~II] line ratios in $10^4$~K, electron density
                   from the [S~II] line ratio in units of 1000 cm$^{-3}$. 
                   Values marked with `A' are
                   default values assumed in absence of the line ratio; densities marked
                   with `H' are lower limits because of the saturation of the [S~II] ratio 
                   for high densities, densities marked with 'C' are from the [Cl~III] lines} 
   \label{t:plasma}
      {\scriptsize 
   \begin{tabular}{lllllllllr}
   \noalign{\smallskip} \hline \noalign{\smallskip}
  PN G & common name  & Spec.Type & Q & EC &$\lg(F({\rm H}\beta))$
                    & c \ \ \ & T(OIII) & T(NII) & n(SII) \\
 \noalign{\smallskip} \hline
   \noalign{\smallskip}
$000.4-01.9$ & M 2-20       &  WC5-6 & A &    4 &$ -12.80$  &  1.71~~~&   ---  &  0.79 &~ 5.40      \\
$002.2-09.4$ & Cn 1-5       &  WO4pe & A &    5 &$ -11.75$  &  0.42~~~&   0.87 &  ---  &~ 4.82      \\
$002.4+05.8$ & NGC 6369     &  WO3   & A &    5 &$ -13.50$  &  1.91~~~&   0.80 &  0.92 &~ 1.74      \\
$003.1+02.9$ & Hb 4         &  WO3   & A &    6 &$ -12.50$  &  1.76~~~&   0.96 &  1.04 &~ 6.71      \\
$004.8-22.7$ & He 2-436     &  WC4   & C &    5 &$ -12.39$  &  0.55~~~&   1.00A&  1.00A& 12.92~C    \\
$004.9+04.9$ & M 1-25       &  WC4   & A &    4 &$ -12.24$  &  1.46~~~&   0.79 &  0.81 &~ 8.00      \\
$006.0-03.6$ & M 2-31       &  WC4   & A &    5 &$ -12.50$  &  1.29~~~&   0.93 &  1.08 &~ 5.31      \\
$006.4+02.0$ & M 1-31       &  wels  & A &    5 &$ -12.73$  &  2.00~~~&   ---  &  0.99 & 11.70      \\
   \noalign{\smallskip}
$006.8+04.1$ & M 3-15       &  WC4   & A &    5 &$ -12.89$  &  2.08~~~&   1.10 &  1.11 &~ 5.56      \\
$009.4-05.0$ & NGC 6629     &  wels  & A &    5 &$ -12.16$  &  0.92~~~&   0.86 &  ---  &~ 2.13      \\
$010.8-01.8$ & NGC 6578     &  wels  & A &    5 &$ -12.49$  &  1.35~~~&   0.83 &  0.88 &~ 7.46      \\
$011.7-00.6$ & NGC 6567     &  wels  & A &    5 &$ -11.47$  &  0.70~~~&   1.09 &  1.09 & 10.46      \\
$011.9+04.2$ & M 1-32       &  WO4pe & A &    4 &$ -12.53$  &  1.30~~~&   1.09 &  0.86 &~ 9.25      \\
$012.2+04.9$ & PM 1-188     &  WC10  & C &$<$ 2 &$ -14.53$  &  1.36~~~&   1.00A&  1.00A&~ 2.29      \\
$016.4-01.9$ & M 1-46       &  wels  & A &$<$ 2 &$ -12.33$  &  1.07~~~&   ---  &  0.69 &~ 2.83      \\
$017.9-04.8$ & M 3-30       &  WO1   & C &    7 &$ -14.07$  &  0.52~~~&   1.00A&  1.00A&~ 3.40      \\
   \noalign{\smallskip}
$019.4-05.3$ & M 1-61       &  wels  & A &    5 &$ -11.74$  &  1.71~~~&   0.93 &  ---  & 16.41      \\
$019.7-04.5$ & M 1-60       &  WC4   & A &    5 &$ -12.67$  &  1.52~~~&   0.86 &  1.00 &~ 6.93      \\
$020.9-01.1$ & M 1-51       &  WO4pe & A &    5 &$ -13.92$  &  3.37~~~&   ---  &  0.88 &~ 7.73      \\
$027.6+04.2$ & M 2-43       &  WC7-8 & C &    4 &$ -12.79$  &  2.67~~~&   1.00A&  1.00A& 11.00~C    \\
$029.2-05.9$ & NGC 6751     &  WO4   & A &    5 &$ -13.06$  &  0.50~~~&   1.06 &  ---  &~ 2.27      \\
$034.6+11.8$ & NGC 6572     &  wels  & A &    5 &$ -10.73$  &  0.30~~~&   1.02 &  1.17 & 17.41      \\
$038.2+12.0$ & Cn 3-1       &  wels  & A &$<$ 2 &$ -11.33$  &  0.44~~~&   ---  &  0.75 &~ 6.90      \\
$048.7+01.9$ & He 2-429     &  WC4   & A &    4 &$ -13.36$  &  2.21~~~&   ---  &  0.84 &~ 7.16      \\
   \noalign{\smallskip}
$055.5-00.5$ & M 1-71       &  wels  & A &    5 &$ -12.52$  &  2.18~~~&   0.90 &  1.26 & 12.39      \\
$057.2-08.9$ & NGC 6879     &  wels  & A &    5 &$ -11.70$  &  0.42~~~&   1.03 &  1.26 &~ 4.16      \\
$061.4-09.5$ & NGC 6905     &  WO2   & B &    7 &$ -12.84$  &  0.00~A &   1.15 &  1.17 &~ 0.53      \\
$068.3-02.7$ & He 2-459     &  WC9   & C &$<$ 2 &$ -13.38$  &  2.65~~~&   1.00A&  1.00A& 16.17      \\
$253.9+05.7$ & M 3-6        &  wels  & A &    5 &$ -12.05$  &  0.49~~~&   0.81 &  1.13 &~ 5.21      \\
$258.1-00.3$ & He 2-9       &  wels  & A &    4 &$ -12.71$  &  2.22~~~&   1.01 &  0.98 & 10.85      \\
$274.6+02.1$ & He 2-35      &  wels  & A &    5 &$ -12.29$  &  0.80~~~&   0.87 &  ---  &~ 1.97~C    \\
$278.1-05.9$ & NGC 2867     &  WO2   & A &    7 &$ -11.57$  &  0.58~~~&   1.19 &  1.11 &~ 2.81      \\
   \noalign{\smallskip}
$278.8+04.9$ & PB 6         &  WO1   & A &    7 &$ -13.38$  &  0.57~~~&   1.58 &  1.13 &~ 2.89      \\
$285.4+01.5$ & Pe 1-1       &  WO4   & A &    5 &$ -12.72$  &  2.16~~~&   0.97 &  1.16 & 18.31      \\
$291.3-26.2$ & Vo 1         &  WC10  & C &$<$ 2 &$ -14.15$  &  2.19~~~&   1.00A&  1.00A&~ 5.00~C    \\
$292.4+04.1$ & PB 8         &  WC5-6 & A &    6 &$ -12.01$  &  0.67~~~&   0.65 &  0.88 &~ 3.99      \\
$300.7-02.0$ & He 2-86      &  WC4   & A &    5 &$ -12.62$  &  2.49~~~&   0.87 &  1.01 & 11.90      \\
$307.2-03.4$ & NGC 5189     &  WO1   & A &    5 &$ -13.33$  &  0.44~~~&   1.21 &  0.85 &~ 0.48      \\
$327.1-02.2$ & He 2-142     &  WC9   & B &$<$ 2 &$ -12.34$  &  2.11~~~&   ---  &  0.75 & 20.00~H    \\
$331.3+16.8$ & NGC 5873     &  wels  & A &    7 &$ -11.56$  &  0.37~~~&   1.40 &  ---  &~ 5.19      \\
   \noalign{\smallskip}
$336.2-06.9$ & PC 14        &  WO4   & A &    5 &$ -12.16$  &  0.65~~~&   0.91 &  1.02 &~ 3.05      \\
$337.4+01.6$ & Pe 1-7       &  WC9   & B &$<$ 2 &$ -12.69$  &  2.79~~~&   ---  &  1.64 & 20.00~H    \\
$351.1+04.8$ & M 1-19       &  wels  & A &    4 &$ -12.27$  &  1.24~~~&   0.69 &  0.83 &~ 5.49      \\
$355.2-02.5$ & H 1-29       &  WC4   & A &    5 &$ -12.81$  &  1.61~~~&   0.93 &  1.04 &~ 7.30      \\
$355.9-04.2$ & M 1-30       &  wels  & A &    3 &$ -12.24$  &  1.01~~~&   0.71 &  0.67 &~ 4.93      \\
$356.7-04.8$ & H 1-41       &  wels  & A &    7 &$ -12.67$  &  0.65~~~&   1.01 &  1.00 &~ 1.17      \\
$357.1+03.6$ & M 3-7        &  wels  & A &    4 &$ -12.80$  &  1.83~~~&   ---  &  0.78 &~ 4.67      \\
$358.3-21.6$ & IC 1297      &  WO3   & A &    7 &$ -11.62$  &  0.18~~~&   1.03 &  1.01 &~ 2.80      \\
   \noalign{\smallskip} \hline
   \end{tabular}
   }
\end{table*}

\section{Chemical compositions of the nebulae}

  With the electronic temperatures and densities obtained from the analysis of 
  line ratios described above, the emissivities of all lines can be computed
  and thus from the observed dereddened intensities the ionic abundances relative 
  to H$^{+}$ are deduced. Whenever both [N~II] and [O~III] electron temperatures are
  available, we use the [N~II] temperature for the low ionization species (N$^{+}$, 
  O$^{+}$, S$^{+}$, S$^{++}$) and the [O~III] temperature for the higher species 
  (O$^{++}$, Ne$^{++}$, Ar$^{++}, $Ar$^{3+}$). To correct for unseen stages 
  of ionization, the usual empirical correction factors (ICF) are applied 
  (see Aller 1984). Table 3 (on-line version), presents the derived abundances of the 
  48 nebulae. All abundances are expressed in the usual logarithmic form of 
  12 + lg($n$(X)/$n$(H)).\\

  For helium, the abundances of $\rm He^{+}$ and $\rm He^{++}$ are derived
  from the He~I and He~II recombination lines. The emissivities of
  the He~I lines are corrected for self-absorption and collisional
  excitation. In low excitation nebulae there could be present an appreciable
  amount of unobservable neutral helium. Therefore the He
  abundances in low excitation nebulae (EC 4 and less) are only lower
  limits and are marked with a colon in Table 3 (on-line version).
  For the sulphur ICF, we use the recipe of Samland et al. (1992), 
  while for chlorine we apply a simple formula that gave a reasonable 
  approximation to results from photoionization models: 
  $$
      {{\rm Cl}\over {\rm H}} \approx {{\rm Cl^{++}}\over {\rm H^+}}
                \left({{\rm He}\over {\rm He^+}}\right)^2
  $$

\subsection{Elements synthesized in progenitor stars}

\begin{figure}
    \includegraphics[trim = 70 70 70 70, scale=0.35, angle=270]{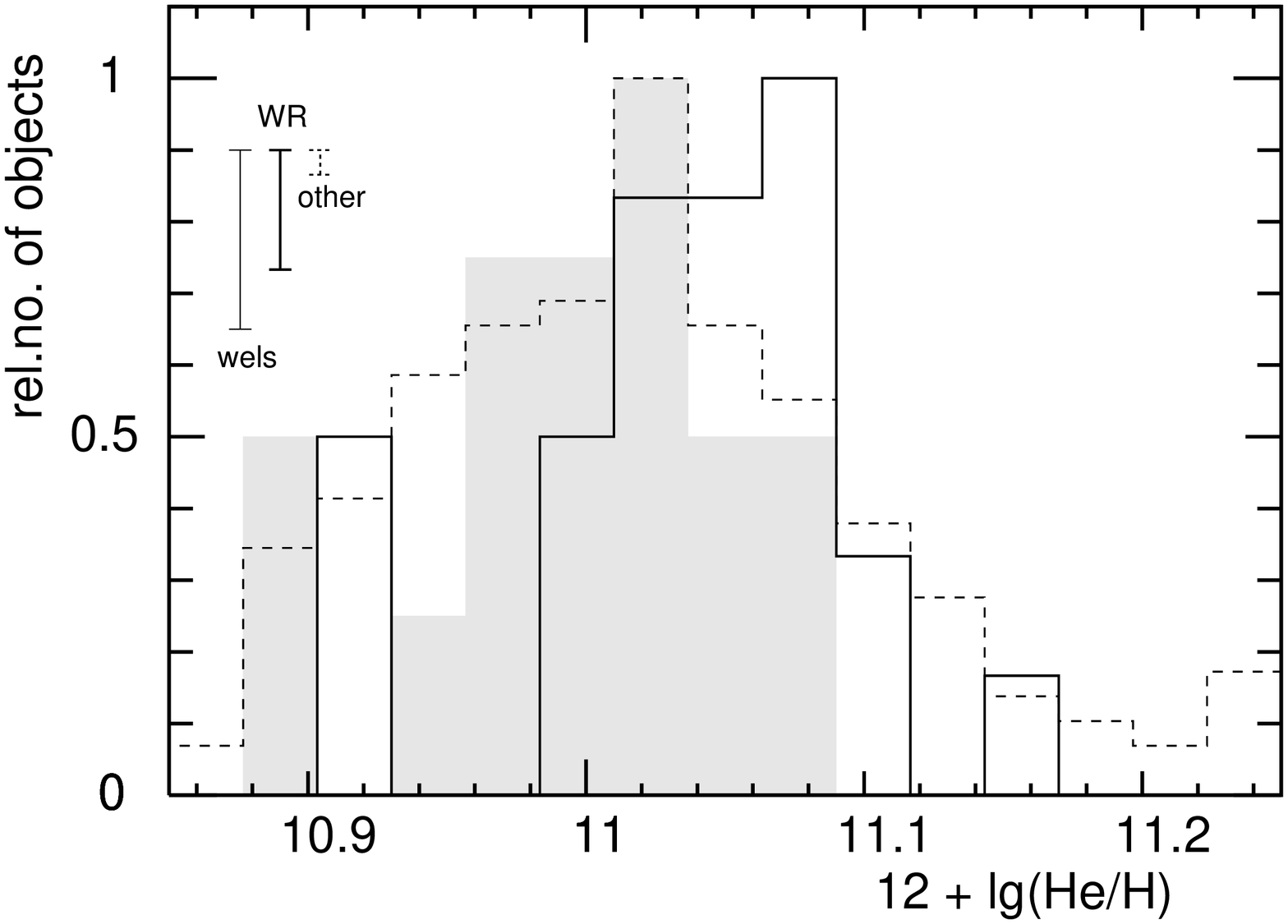} 
      \caption[]{Distribution of helium abundances among [WR] PN
                 (full line),  {\it wels} objects (shaded area), 
                 and nebulae around `normal' central stars (dashed line). 
                 The vertical error 
                 bars indicate the contribution by a single object.}
      \label{f:he}  
   \end{figure}

  Helium and nitrogen are elements that are produced in more massive 
  progenitor stars of PN. Hence abundances of these elements is expected 
  to be an indicator of the mass of the progenitor star. Comparing the 
  histogram of helium abundances (Fig. \ref{f:he}), we find no evidence 
  for any significant difference between the different types.
  Using only data without colon and quality code Q=A analyses
  from 11 [WR] PN, 14 {\it wels}, and 87 `normal' PN, we  obtain respective
  average values of $11.05 \pm  0.02$, $10.97 \pm  0.03$, and $11.02 \pm 0.01$,
  which places the [WR] PN above the solar value and the normal PN, but the
  {\it wels} have lower helium abundances. The dispersions are $0.05 \pm  0.01$,
  $0.10 \pm  0.02$, and  $0.10 \pm  0.01$, respectively.
 
\begin{figure}[!ht]
      \includegraphics[trim = 70 70 70 70, scale=0.35, angle=270]{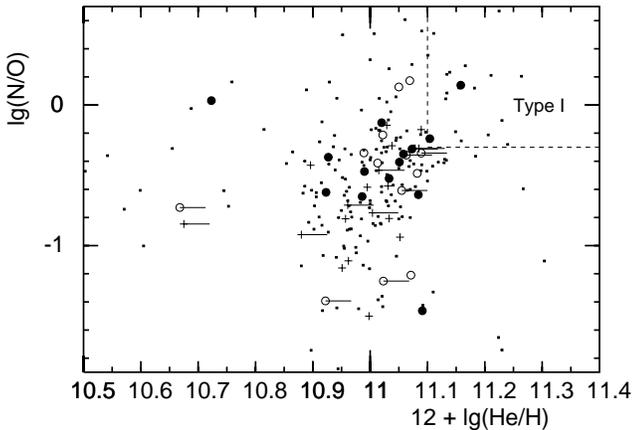} 
       \caption[]{Relation between helium abundances and N/O abundance ratios
                  for WR, {\it wels} and `normal' PN. The symbols are as in Fig. 
                  \ref{f:diffte}. Short bars indicate a lower limit to
                  the He abundance in WRPN of excitation class below 5}
     \label{f:nohe}   
    \end{figure}

  Likewise, the relation between helium abundance and the N/O abundance ratio
  exhibits no clear difference between the two types of nebulae. Neither is obvious
  any difference between objects with [WO], [WC], or {\it wels} central stars, as
  depicted in Fig. \ref{f:nohe}. Two objects fullfill the criterion 
  of Peimbert \& Torres-Peimbert (1983) for Type I nebulae:
  the [WO] types PNG~$002.2-09.4$ and PNG~$278.8+04.9$, and with the {\it wels} 
  PNG~$006.4+02.0$ doing nearly so.

  The average helium abundances for all 
  types, presented in Table 4, are solar without any significant differences. The nitrogen abundance in [WC] 
  and [WO] is somewhat enhanced with respect to the Sun, but {\it wels} and `normal' PN 
  have solar values. The N/O abundance ratio found in both [WC] and [WO] type nebulae 
  is about thrice solar, but in `normal' nebulae it is not more than twice solar.
  The {\it wels} objects have nearly a solar ratio. One also notes that the helium 
  abundance is somewhat lower, by about twice the standard error. It thus is 
  tempting to identify {\it wels} nebulae as coming from less massive progenitor 
  stars than the other types of PN.

  \begin{table*}
  {Table 4. Average abundances for the PN of each type, along with the 
\               dispersion.
                Only values without a colon in Table 3 from class `A' analyses 
                are taken into account; 
                numbers of objects used are given in parenthesis.
                The values for the non-WR sample come from our reanalysis of the
                published data. The solar system abundances are from Anders \& Grevesse (1989)}\\ 
\centering
     \begin{tabular}{llllll}
     \noalign{\smallskip} \hline \noalign{\smallskip}
      Element & [WC] & [WO] & $wels$ & non-WR & solar \\
    \noalign{\smallskip} \hline
     \noalign{\smallskip}
      He & $11.06\pm 0.03$ (3)& $11.06\pm 0.06$ (6)& $10.97\pm 0.10$ (14)& $11.02\pm 0.10$  (87) & 10.99 \\
      N   & $8.43\pm 0.28$ (9)&  $8.37\pm 0.23$ (12)&  $7.94\pm 0.38$ (18)&  $8.13\pm 0.55$  (144)  &  8.05 \\
      O   & $8.70\pm 0.13$ (9)&  $8.72\pm 0.14$ (12)&  $8.64\pm 0.15$ (18)&  $8.57\pm 0.26$  (155) &   8.93 \\
      Ne  & $7.93\pm 0.24$ (8)&  $8.10\pm 0.25$ (12)&  $7.94\pm 0.16$ (15)&  $7.92\pm 0.30$  (88) &   8.09 \\
      S   & $7.05\pm 0.20$ (9)&  $7.06\pm 0.25$ (12)&  $6.79\pm 0.25$ (17)&  $6.78\pm 0.30$  (109)  &  7.21 \\
      Ar  & $6.58\pm 0.17$ (7)&  $6.40\pm 0.25$ (9)&  $6.34\pm 0.32$ (12)&  $6.22\pm 0.28$  (80)  &  6.56 \\
     \noalign{\smallskip} 
      N/O & $-0.27\pm 0.25$ (9) & $-0.35\pm 0.25$ (12) & $-0.70\pm 0.35$ (18)& $-0.44\pm 0.48$ (143) & $-0.88$ \\
     \noalign{\smallskip} \hline
     \end{tabular}
  \end{table*}

\subsection{The other elements}

  Oxygen, neon, sulphur, and argon are synthesized in massive stars, and their abundances 
  are not altered by the nucleosynthesis in the PN progenitor stars. 
   The average oxygen abundances, presented in Table 4, among [WC] and [WO] PN
  are slightly above the values found in {\it wels} and `normal' PN, all of which are 
  significantly lower (about 0.3 dex) than the solar system abundance. Neon is close to
  solar values in all types. Sulphur is about 0.3 dex higher in [WO] and [WC] than in
  {\it wels} and `normal' PN. In all PN sulphur appears to be lower than the solar value; 
  however, one has to keep in mind that the empirical ICF for sulphur is less accurate 
  than for nitrogen or neon. Thus we do not want to exclude a systematic tendency for 
  underestimating the sulphur abundances. Within the error bars, argon is also
  solar in all types of nebulae. Therefore, the chemical compositions of planetary nebulae 
  with central stars of types [WC], [WO], and {\it wels} appear to differ somewhat from
  those of nebulae with `normal' central stars,  with nearly solar system values. 

\begin{figure}
    \includegraphics[trim = 70 70 70 70, scale=0.35, angle=270]{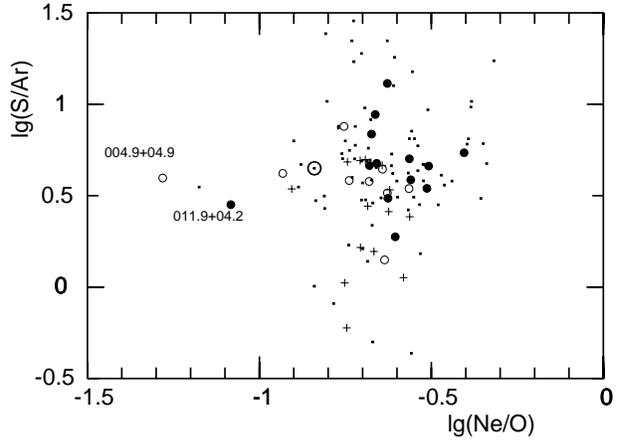} 
     \caption[]{Relation between the abundance ratios Ne/O and S/Ar. The symbols are as in
                Fig. \ref{f:diffte}. The solar symbol indicates solar values 
               ($\lg({\rm Ne/O}) = -0.84$ and $\lg({\rm S/Ar}) = 0.65$) }
     \label{f:neosar}  
  \end{figure}

   The abundance ratios Ne/O and S/Ar, shown in Fig. \ref{f:neosar}, of either type of PN are 
  rather close to the ratios found in the Sun. As already seen from Table 4, all PN cluster
  around the solar system value for S/Ar ratio, but at higher Ne/O ratio. The distributions 
  of the [WC] and [WO] nebulae are quite similar to those of the `normal' PN. However,
  a slight difference can be noted, in that the [WO] tend to have higher Ne/O and S/Ar ratios, 
  while among the [WC] both ratios are smaller. The {\it wels} also show a preference for having 
  lower S/Ar ratios. Because this pattern is already present, if one considers only nebulae
  of the same excitation (e.g. 5), these differences appear to be real, and not to be artifacts
  caused by the use of ICF, for example.

  There are two remarkable outliers, PNG~$004.9+04.9$ (one of the three nuclei of [WC~5-6] type) 
  and PNG~$011.9+04.2$ (one of the four nuclei of type [WO~4p] which show very high velocity winds), 
  which also stand out in diagrams of the other combinations of abundance ratios; neither object 
  shows any obvious flaw in the observational material or analysis. In these objects, PSM01 had 
  found unusually low Ne/O abundance ratios of 0.00343 and 0.00371, respectively. Our values 
  (0.052 and 0.083) are higher, but still substantially lower than the solar value of 0.16. The 
  other non-WR object in the vicinity is PNG~$325.4-04.0$ observed by Kingsburgh \& Barlow (1994).

\section{Other global relations}

\subsection{Diagnostic diagram}

\begin{figure}[!ht]
      \includegraphics[trim = 70 70 70 70, scale=0.35, angle=270]{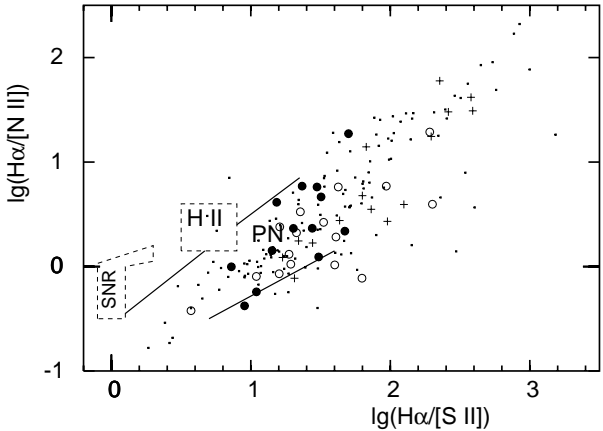} 
       \caption[]{The intensity ratios H$\alpha$/[N~II] and H$\alpha$/[S~II] of our objects,
                  in relation to the regions identified by Cant\'o (1981). The symbols 
                  are as in Fig. \ref{f:diffte}.}
     \label{f:canto}   
   \end{figure}

   In the diagnostic diagram of the intensity ratios H$\alpha$/[N~II] and H$\alpha$/[S~II]
   (Fig. \ref{f:canto}, after Cant\'o 1981 and Corradi et al. 1997), the nebulae of our sample 
   cover the region occupied by planetary nebulae. Except for some preference of the [WO]  
   objects to do not extend to high H$\alpha$/[S~II] and H$\alpha$/[N~II] ratios, no specific 
   region for [WR] nebulae can be distinguished.\\

\subsection{Dereddened flux and angular diameter}

\begin{figure}[!ht]
      \includegraphics[trim = 70 70 70 70, scale=0.35, angle=270]{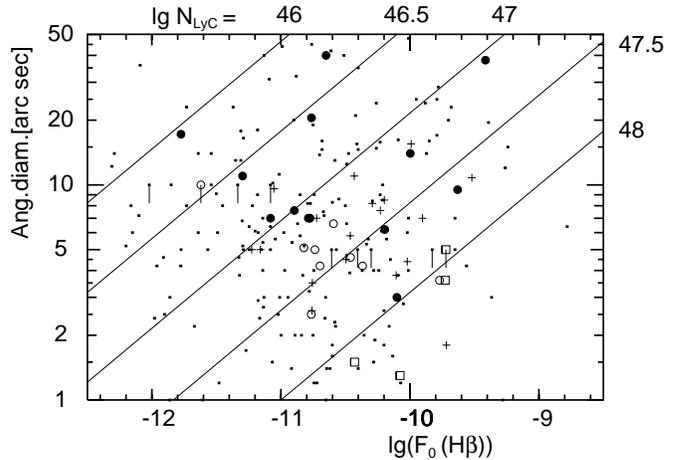} 
       \caption[]{Dereddened H$\beta$ fluxes and angular diameters of nebulae with central 
                  stars of the various types:
		  [WO] (filled circles),
               [WC] 5-6 and 4 (open circles), [WC] 10, 9, and 7-8 (open squares), 
	       {\it wels} (plus-signs), and any other (small dots).
	       Short vertical lines
                  mark objects whose diameter is only an upper limit.
                  The lines of constant number of ionizing photons (in photons s$^{-1}$)
                  are computed for ionization bounded nebulae of 0.25 solar masses}
     \label{f:fludia}   
   \end{figure}

   The diagram of dereddened H$\beta$ flux and angular diameter allows some inferences
   on the number of ionizing photons from the central star: In an ionization bounded 
   Str\"omgren sphere of radius $R$, the dereddened H$\beta$ flux at a distance $d$ depends 
   on the number of ionizing photons as
   \begin{equation}
           {\cal N}_{LyC} = {4\pi\over 3} \alpha_B R^3 n^2 = 
           {\alpha_B\over \alpha_{\rm eff}} \cdot 4\pi d^2 \cdot F_0({\rm H}\beta )
   \end{equation}
   with the coefficients $\alpha_B$ for all recombinations to excited levels and
   $\alpha_{/rm eff}$ for leading to the emission of H$\beta$ photons.
   Using the relation $R/d = \sin(D) = D$ between angular diameter $D$ and $R$, one obtains
   \begin{equation}
          F_0 d^2 \propto F_0/D^2 \cdot {\cal N}_{LyC}^{2/3}\cdot n^{-4/3}
   \end{equation}
   and demanding that the ionized mass $M_{\rm ion} =  4\pi n m R^3/3$ remains constant, 
   one gets 
   \begin{equation}
         F_0 \propto D^2 \cdot {\cal N}_{LyC}^{5/3} \cdot M_{\rm ion}^{-4/3}
   \end{equation}
    From the catalog of Acker et al. (1992) we take the optical diameter or  -- if that
   datum is uncertain, an upper limit, or the object is described as `stellar' -- the radio
   diameter. Figure \ref{f:fludia} shows that `normal' PN have a wide distribution in both
   diameter and number of ionizing photons. The same characteristic is also found among 
   the [WO] types, as far as one can judge from the limited number of 
   objects. However, the [WC] nebulae have a strong tendency to be smaller: except 
   PNG $004.8-22.7$ for which the optical diameter of 10 arcsec is only an upper limit,
   all nebulae are smaller than 6.6 arcsec (PNG $000.4-01.9$). Late-type [WC] 10, 
   9, and 7-8 objects (shown as open squares) are smaller than about 5 arcsec, have
   high H$\beta$ fluxes, and high number of ionizing photons, about 
   $\lg{\cal N}_{LyC} >  48$. All this ties in well with the notion that [WC] objects are 
   young and compact nebulae. 
   The {\it wels} have a fairly wide range in diameters and in ionizing photon number, 
   although one may note that none is present with diameters larger than about 15 arcsec
   and $\lg{\cal N}_{LyC} <  47.3$.   
   One also has to keep in mind that these numbers depend on the ionized mass of the nebula,
   which might vary among the types.\\

\subsection{Infrared colours}

\begin{figure}[ht] 
    \begin{center}
      \includegraphics[trim = 70 70 70 70, scale=0.35, angle=270]{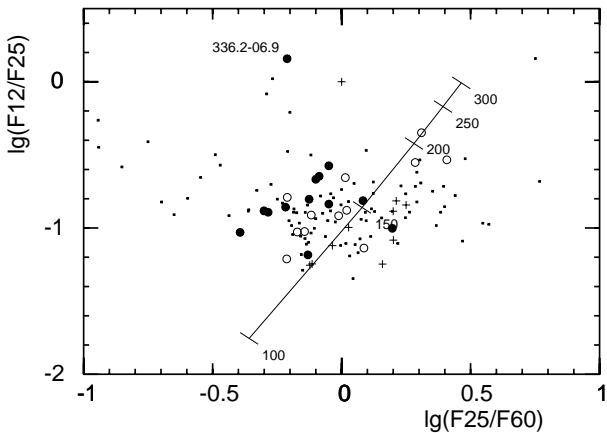} 
     \caption[]{Two-colour diagram of IRAS fluxes at 12, 25, and 60 $\mu$m.
                Symbols have the same meaning as in Fig. \ref{f:diffte}.
                Only high quality data is included. The line with ticks indicates the 
                colours of black-body spectra between 100 and 300 K}
     \label{f:irasa}  
    \end{center}   
  \end{figure}

Planetary nebulae are strong infrared emitters, due to their heated dust: 
as the nebula expands, the dust cools and the dust colours redden. Fig. \ref{f:irasa} 
shows the two-colour diagram for our samples.  The [WR] objects are found mainly 
towards the left in the diagram and the {\it wels} 
are shifted to bluer colours, below the cooling line. 

By using the dust model of Siebenmorgen et al. (1994), the IRAS fluxes can be 
calculated at any time following the nebular expansion, both for carbon-rich and 
oxygen-rich dust (see Acker et al. 1996a, and Gesicki et al. 2006). 
From the Fig. \ref{f:irasa} of this paper and the Fig. 7 of Gesicki et al. (2006), 
the [WR] stars are 
displayed along evolutionary progression of carbon-rich nebular dust, from cool stars 
with bluer IRAS colours (like those of post-AGB stars) to hot stars with redder IRAS 
colours. But the {\it wels} sample appears to be shifted below the oxygen rich track. 
Note that most of the {\it wels} were classified as O-rich stars by M\'endez (1991).

On the other hand, Gesicki et al. (2006) show that about 65\% of the CSPNe have 
good IRAS detections, with different detection rates: respectively 52\%, 53\%, and 
88\% for `normal', {\it wels} and [WR] PNe, the latest being on average much stronger 
IRAS emitters. If one considers 25 and 60 $\mu$m only, the detection rates become 
89\%, 86\% and 100\%, i.e. the difference is rather reduced, which indicates that 
[WR] PNe (essentially [WC] PNe) are stronger emitter at 12 $\mu$m. This is indicative 
of small dust grains, which is not found in {\it wels}.

The [WR] evolution seems marked by the presence of nebular turbulence, almost universal 
for [WR] PNe, common for {\it wels} but rare for `normal' PNe (91\%, 24\%, and 7\%, 
respectively), a difference which shows that the `normal' PNe are not closely related 
to the [WR] stars, as the different dynamics of the nebulae takes time to build up 
(Acker et al. 2002, Acker \& Neiner 2003, Gesicki et al. 2006). Given the differences 
in IRAS colours, chemistry and carbon surface abundances of the [WC] stars, a sequence 
can be considered where they evolve too slowly to reach [WO] types while still 
surrounded by a bright nebula. Gesicki et al. (2006) show that strong correlation 
appears between enhanced 12$\mu$m flux and turbulence for the [WC] PNe, leading to 
the formation of small grains. But this is not the case for the {\it wels} which show 
almost identical colours regardless of turbulence. The expansion is a strong function 
of metallicity, therefore the expansion of the nebulae around 
O-rich stars grows slower than around C-rich stars (Acker et al. 1996a), 
and shows lower 
dust abundance in the AGB wind (Wood et al. 1992). The very blue [WO~4] nebula PC~14 (PNG~$336.2-06.9$) 
is perhaps more compact than other [WO~4] PNe, with a lower expansion related to a lower 
metallicity. We also note that many [WC] stars show a unique feature in their nebular 
chemistry, with both PAHs and oxygen-rich dust and gas being present (Zijlstra 2001). 
This mixed chemistry is not seen among either [WR] PNe or any other group of PNe. 

All these aspects confirm the analysis by Tylenda et al. (1993), who argue that the 
[WR] and the {\it wels} form two distinct groups: {\it wels} show very weak emission 
lines, and a different chemistry in the line forming wind, the N/C ratio being higher 
in {\it wels} than in [WC] (Acker at al. 1996a). 
The [WR] PNe seem to be more compact than the nebulae around 
{\it wels}, which indicates a different mass-loss history. The progenitor properties 
(age, metallicity, stellar activity) may play the major r\^ole in the formation of a 
{\it wels} or a [WC] star.

\begin{figure}[ht] 
    \includegraphics[trim = 70 70 70 70, scale=0.35, angle=270]{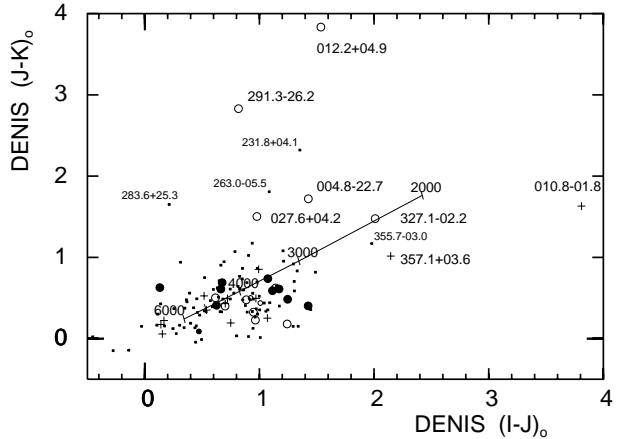}
     \caption[]{The two-colour diagram of I-J vs J-K from DENIS data.
                The line with ticks indicates colours from blackbody
                with temperatures from 3000 to 6000 K. 
                The symbols are as in Fig. \ref{f:diffte}}
     \label{f:denis}  
  \end{figure}

  From the DENIS database at CDS we extracted the infrared colours for our objects
  (10 [WO], 14 [WC], 13 {\it wels}, and 94 non-WR nebulae). In the two colour diagram 
  (Fig. \ref{f:denis}) of the I(0.82$\mu$m), J(1.25$\mu$m) and K(2.15$\mu$m) bands, 
  most nebulae of all types cluster near colour temperatures of 4000 K. While none 
  of the [WO] objects is found elsewhere, several [WC] nebulae are found at 
  high (J-K)$_0$ colour indices: foremost PNG~$012.2+04.9$ and PNG~$291.3-26.2$
    which have been suspected for the presence of hot dust (G\'orny et al. 2001), and
    also PNG~$004.8-22.7$, PNG~$027.6+04.2$, and PNG~$327.1-02.2$. 
  
  Likewise, two {\it wels} are found with high (I-J)$_0$ values: PNG~$010.8-01.8$ and 
  PNG~$357.1+03.6$. Inspection of the DENIS data reveals that none of these objects 
  has magnitude errors in excess of the other objects. Neither do we find correlations 
  with other nebular properties, such as H$\beta$ flux, extinction, or angular diameter.

\section{Relations between central star and nebula}

\begin{figure}
    \includegraphics[trim = 70 70 70 70, scale=0.35, angle=270]{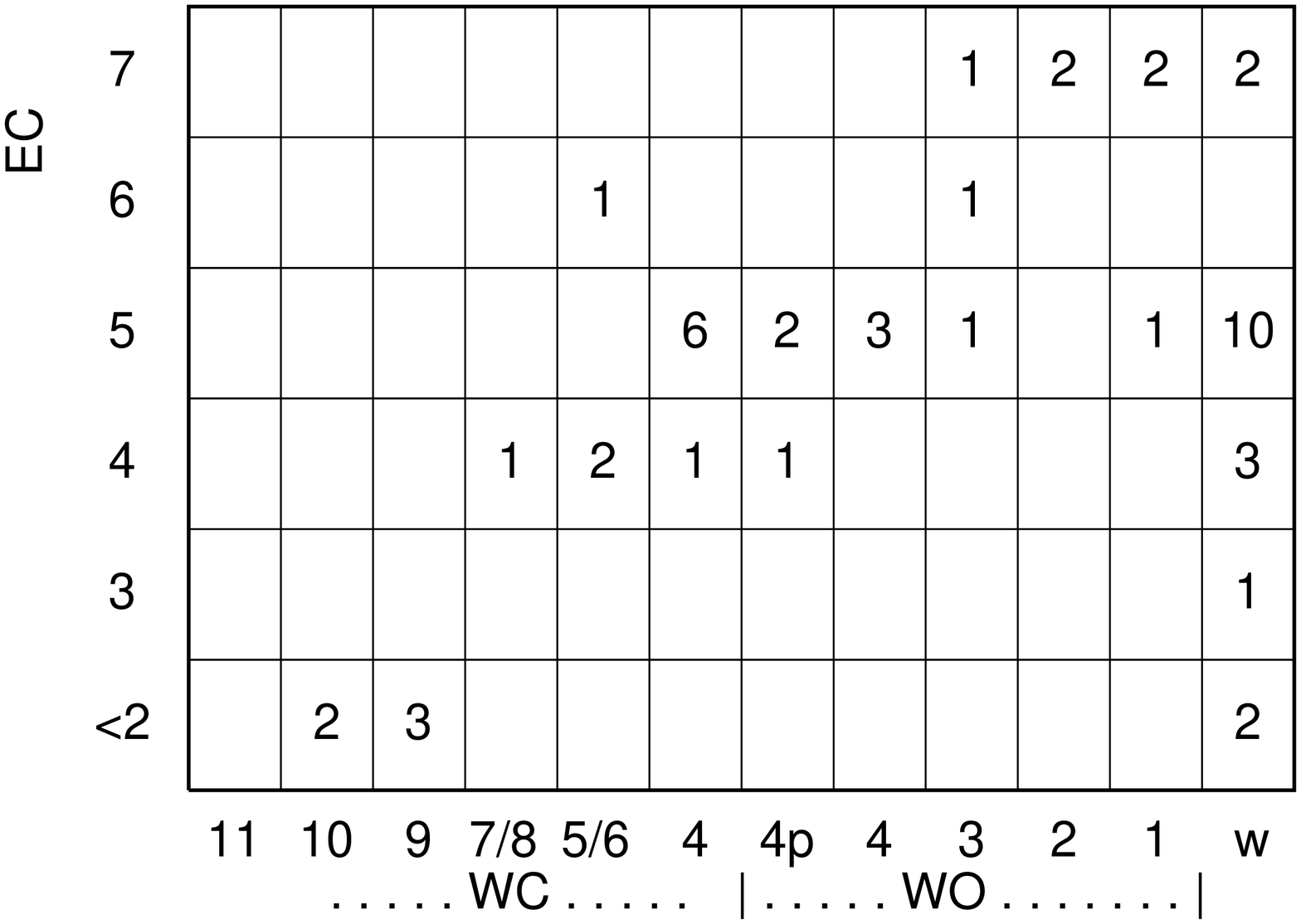} 
     \caption[]{Distribution of nebulae among central star spectral types and excitation
                class of the nebula. The column marked `w' refers to {\it wels} objects}
     \label{f:stec}  
  \end{figure}

  To explore the possibility of an evolutionary sequence from [WC~11] to [WO~1], 
  we investigate the nebular properties as a function of the spectral type of their 
  central stars. Since the excitation class is a measure of the temperature
  of the ionizing source, it is expected to be linked to the temperature
  of the central star. Figure \ref{f:stec} indicates a clear relation
  between spectral type and excitation class, in the sense that [WC] central stars
  are found in low excitation nebulae, while the early [WO] stars are surrounded by 
  nebulae of highest excitation. The {\it wels} central stars are found in objects of
  any level of excitation. Given the small sample, we regard the large number 
  with EC = 5 as an observational selection effect.   

\begin{figure}[ht]
    \includegraphics[trim = 70 70 70 70, scale=0.35, angle=270]{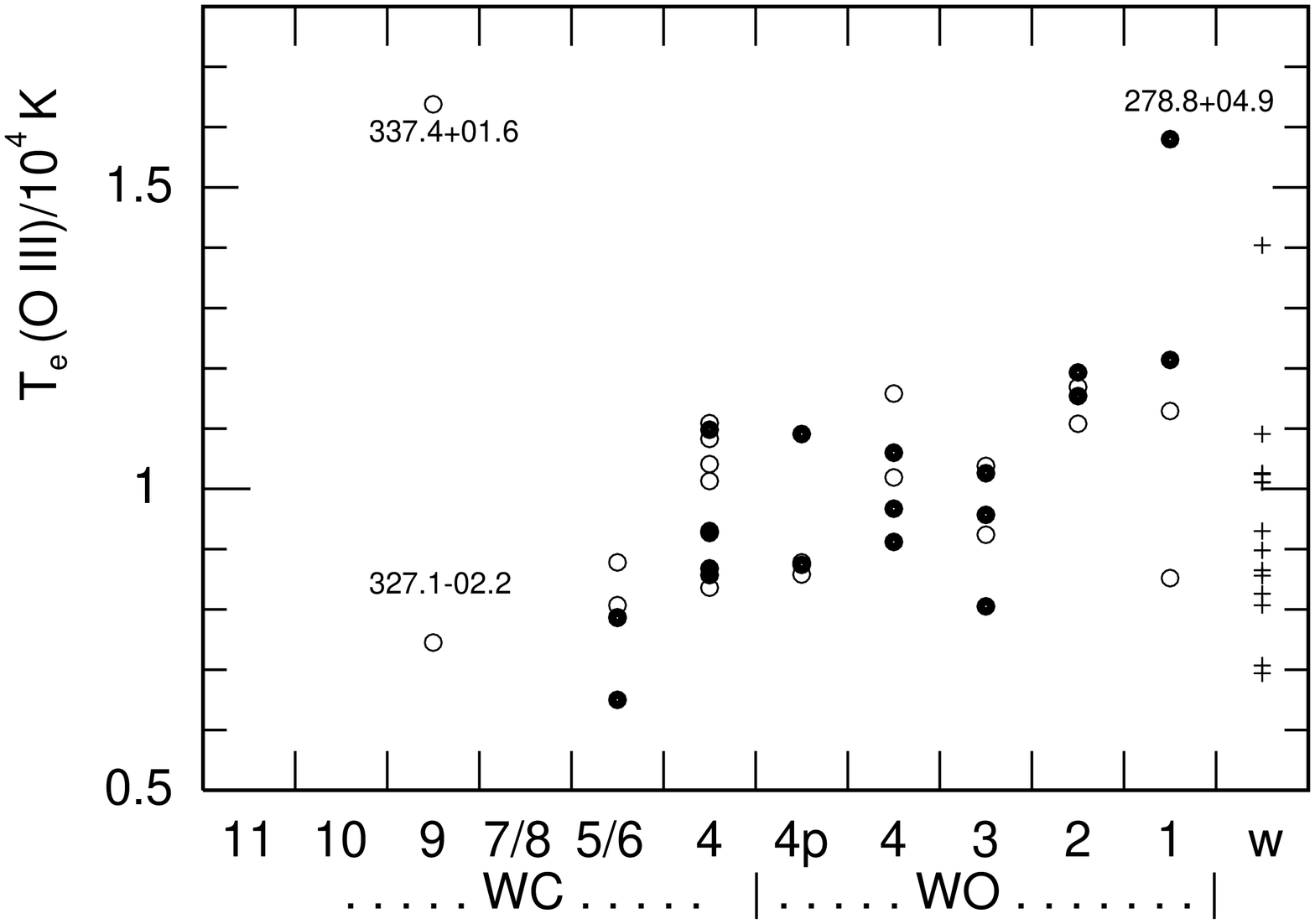} 
    \caption[]{Relation between spectral types and electron temperature from [O~III] (filled
               circles) and from [N~II] (open circles). {\it wels} are marked by a cross}
     \label{f:stto3v2}  
  \end{figure}

  This is also evident in the electron temperatures, depicted in Fig. \ref{f:stto3v2}. 
  Unfortunately, there are no measurements of the [O~III] temperature
  in the low excitation nebulae, and the few reliable [N~II] temperatures do not
  allow to follow this correlation among these types of objects.

\begin{figure}[ht]
      \includegraphics[trim = 70 70 70 70, scale=0.35, angle=270]{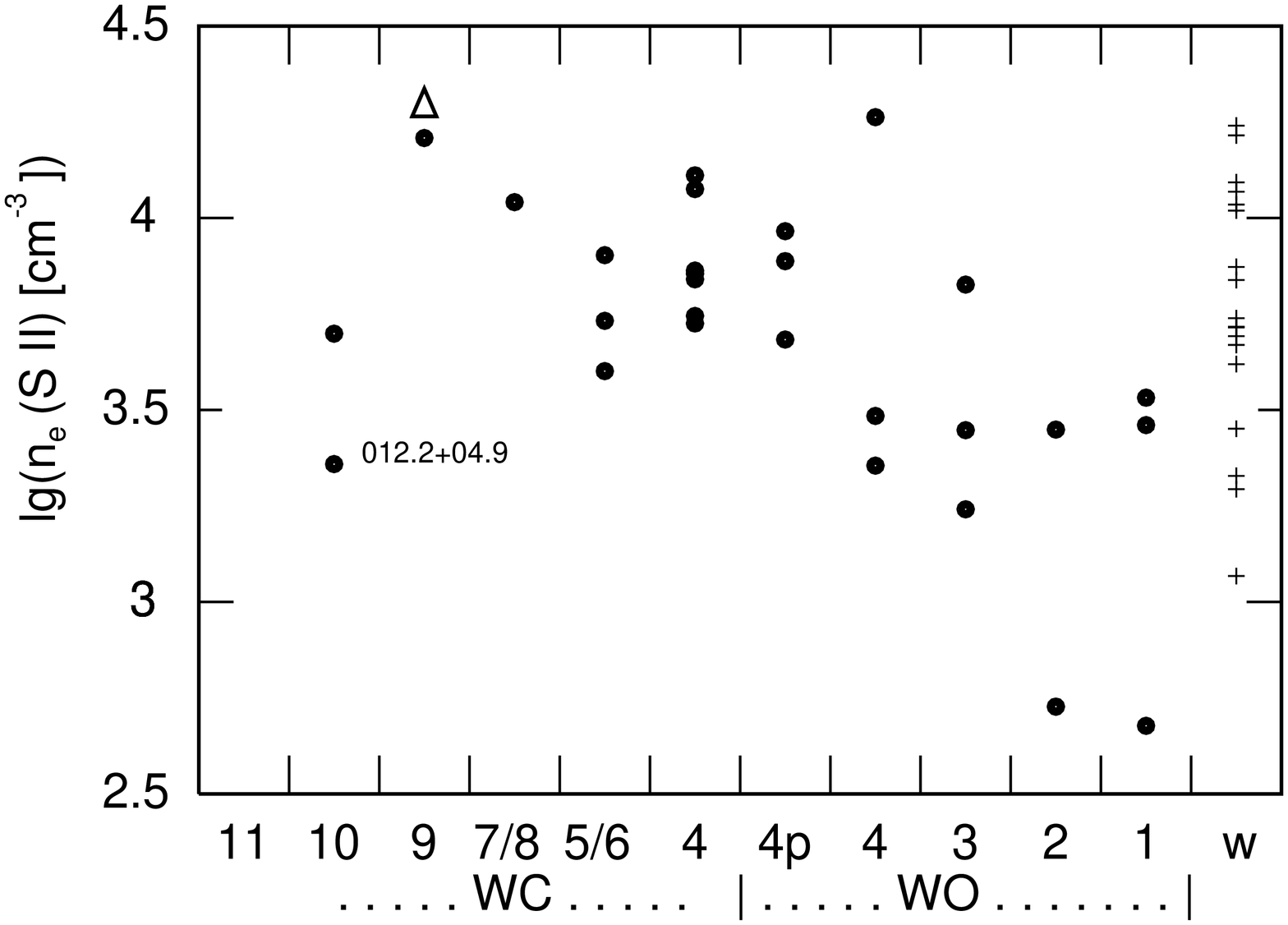}
      \caption[]{Relation between spectral types and electron density from [S~II]. 
                 `$\Delta$' are the limit of density determination (marked 20.00H in table 
                 \ref{t:plasma}). {\it wels} are marked by a cross}
     \label{f:stdens}  
   \end{figure}

  Fig. \ref{f:stdens} shows that the electron density decreases along the sequence from 
  cooler [WC~10] to hot [WO~1]. This ties in well with the notion of an evolutionary sequence: 
  as the density decreases due to the nebular expansion, the central star becomes hotter.
  In contrast, the {\it wels} objects display a wide range of densities, and 
  cannot be incorporated into this sequence. 

  All these results support an evolutionary sequence from [WC~11] central stars which 
  are cool and surrounded by dense, low excitation nebulae towards hot [WO~1] stars
  with low density, high excitation nebulae.

  The {\it wels} objects evidently belong to a different class of planetaries which
  do not appear to be evolutionarily related to the [WR] type nebulae.

\section{Conclusions}

  Nebular spectra of 48 PN around central stars of [WR] and {\it wels} spectral type
  are analyzed for plasma properties and chemical compositions. Comparison of
  the results for [WC], [WO] and {\it wels} objects with the properties of `normal'
  non-WR type objects confirm that the nebulae of either group are very similar:
   \begin{itemize}
      \item they remain indistinguishable in Canto's diagnostic diagram 
      \item the relation of electron temperatures from [N~II] and [O~III] lines
            show the same trend and range
      \item the average [O~III] temperature in [WR] PN and {\it wels} objects is about 
            3000~K lower than in non-WR nebulae
      \item they have the same average helium abundances, and abundance
            distributions. {\it wels} have slightly lower helium abundances
      \item the nitrogen abundances in [WC] and [WO] nebulae are somewhat enhanced
            with respect to the Sun, while {\it wels} and non-WR PN have solar 
            values
      \item the N/O abundance ratio in both [WC] and [WO] PN  is about thrice solar, 
            somewhat higher than in non-WR (less than twice solar). But {\it wels} 
            objects have a much lower ratio, of nearly solar value
      \item O, Ne, S, Ar abundances are nearly solar in all groups.
      \item $[$WC$]$ and {\it wels} nebulae tend to be smaller than [WO] and non-WR 
            objects
      \item the number of ionizing photons covers the same range and has the same
            average value 
      \item  in the IRAS two-colour diagram, {\it wels} are shifted to bluer
            colours than the other [WR] PN
      \item in the DENIS two-colour diagram some [WC]s are found at high (J-K)$_0$ 
            values while two {\it wels} have high (I-J)$_0$ colour indices.
   \end{itemize}

   With respect to the central star's spectral type, some clear trends are present:
   From [WC~11] to [WO~1], 
   \begin{itemize}
      \item the excitation class rises, hence the temperature of the star's 
            ionizing spectrum
      \item the electron temperature rises
      \item the electron density decreases
   \end{itemize}
   as one would expect for an evolutionary sequence from late to early spectra type, 
   as the star heats up and the nebula expands.
   The {\it wels} nebulae have properties that cover a wide range, and they evidently 
   belong to a separate subclass of PN, and do not appear to be evolutionarily related 
   to the [WR] type nebulae.

   Thus our results corroborate the evolutionary sequence from [WC~11] central stars which 
   are cool and surrounded by dense, low excitation nebulae towards hot [WO~1] stars with 
   low density, high excitation nebulae with embedded cooler dust. 
   On the other hand, there is no evidence for the {\it wels} being linked by evolution 
   to the [WR] PN. Rather they seem to constitute a separate class of objects with 
   a variety of nebular properties. Their lower N/O ratio and the hint of a lower 
   He/H abundance suggests that they might have formed from less massive progenitor 
   stars than the other PN.

\begin{acknowledgements}
   We thank Sophie Durand for a first measurement of the majority 
   of the spectra. We wish to express our thanks to the referee for 
   detailed comments and constructive suggestions.
\end{acknowledgements}

\appendix

%
%
%
%
  \section{Comparison of the methods}
        \label{s:comparemethods}

  We reanalyzed with HOPPLA the dereddened line intensities of the objects 
  in the other works and compare the plasma parameters and abundances with 
  the authors' original values (except for CAK96 and CMAKS00 who also use 
  HOPPLA). 
  
\begin{figure}[!ht]
      \includegraphics[trim = 70 70 70 70, scale=0.35, angle=270]{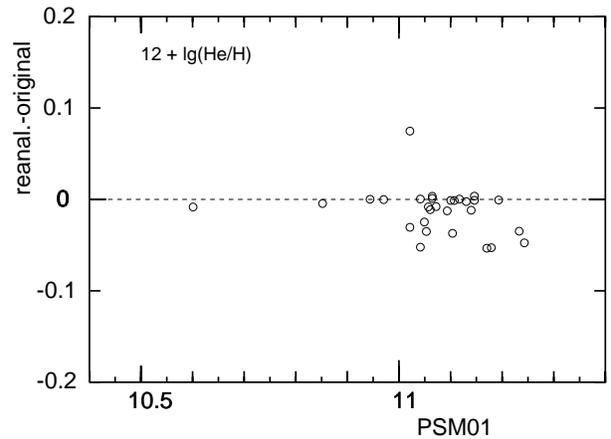} 
       \caption[]{The helium abundances of the PN observed by PSM01
                  compared to the values found by HOPPLA from the same data}
     \label{f:hepena}   
   \end{figure}

\begin{figure}[!ht]
      \includegraphics[trim = 70 70 70 70, scale=0.35, angle=270]{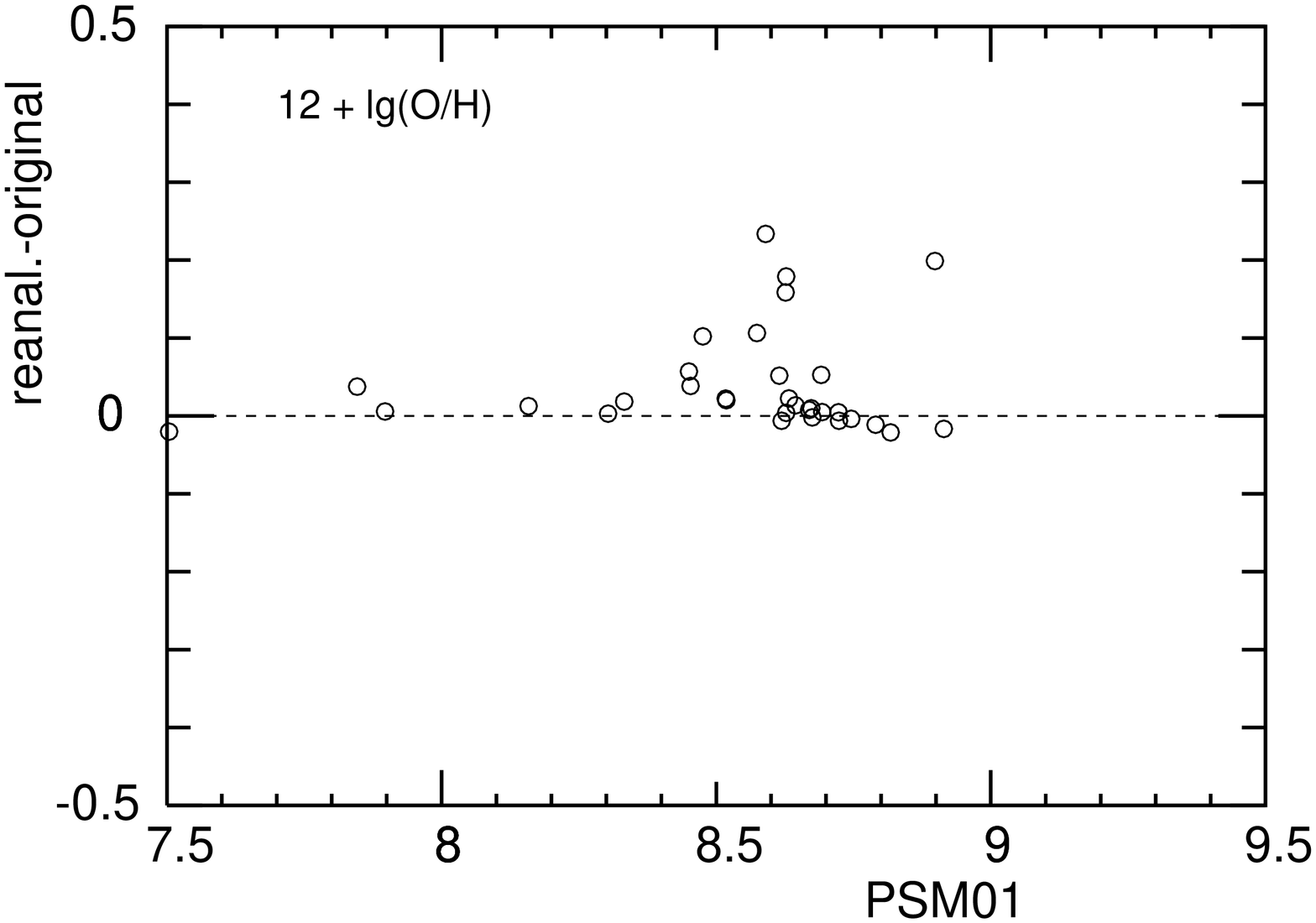} 
       \caption[]{The oxygen abundances of the PN observed by PSM01
                  compared to the values found by HOPPLA from the same data}
     \label{f:opena}   
   \end{figure}

\begin{figure}[!ht]
      \includegraphics[trim = 70 70 70 70, scale=0.35, angle=270]{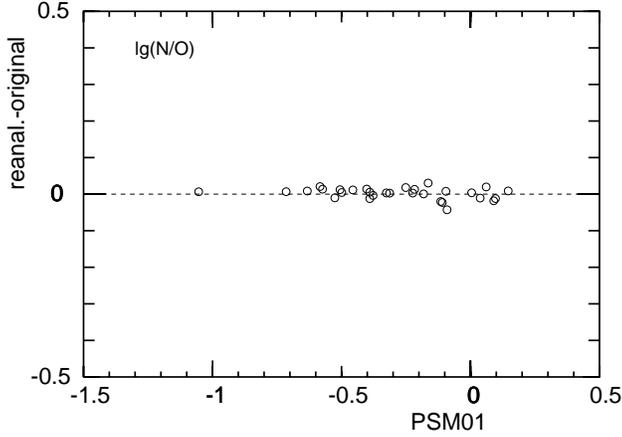} 
       \caption[]{The N/O abundance ratios of the PN observed by PSM01
                  compared to the values found by HOPPLA from the same data}
     \label{f:nopena}   
   \end{figure}

  Since both HOPPLA and PSM01 use nearly the same atomic data, and use rather
  similar strategies, the reanalysis of their data gives almost identical results: 
  both [O~III] and [N~II] electron temperatures are retrieved within 100~K (i.e. 
  within the 3 decimal places of the authors' data), and the [S~II] densities are 
  within 0.05~dex. As shown in Fig. \ref{f:hepena}, the helium abundances mostly 
  agree within 0.01~dex; however, for several high excitation objects (PNG~$017.9-04.8$, 
  PNG~$061.4-09.5$, PNG~$144.5+06.5$, PNG~$161.2-14.8$, PNG~$189.1+19.8$, PNG~$243.3-01.0$, 
  PNG~$278.1-05.9$, PNG~$278.8+04.9$, PNG~$286.3+02.8$) HOPPLA yields a smaller value, 
  by as much as 0.06~dex. None of the objects shows discrepancies in electron
  density or temperature; however, in PNG~$061.4-09.5$ and PNG~$278.1-05.9$ the
  He~I intensity is given as a single digit only (4 and 5), which could mean
  an uncertainty of as much as 20 percent. Another exception is PNG~$356.2-04.4$
  for which a  higher He abundance is found (by 0.08~dex). The origin for
  this discrepancy could not be found.
  The oxygen abundances (Fig. \ref{f:opena}) are recovered within 0.1 dex, 
  with the exceptions of PNG~$061.4-09.5$, PNG~$161.2-14.8$,
  PNG~$189.1+19.8$, PNG~$243.3-01.0$, PNG~$278.8+04.9$, and PNG~$286.3+02.8$ which 
  have values as much as 0.2~dex higher than in PSM01. The origin is
  that in these high excitation objects the oxygen ICF depends on the
  (high) He$^{++}$/He ratio which propagates and amplifies the deviation
  found in the He abundances.
  The maximum deviation in the N/O abundance ratio (Fig. \ref{f:nopena}) 
  is 0.05~dex in PNG~$004.9+04.9$.
  We conclude that both methods give nearly identical results.

  We refrain from showing for the other works the plots of comparing original 
  and re-derived values for all quantities, but present the essential results 
  in Table \ref{t:reanalysis} in numerical form. We summarize the inspection 
  of the plots of key quantities:
  \begin{itemize}
     \item KB94 include lines from the range 1200--3000~\AA\ . Whether
           or not these lines are included in the analysis, has rather little
           impact, but in all plots we show only the results based on the optical
           lines. The [O~III] temperature is recovered 
           within 300~K, with several outliers which cause a non-zero
           mean and a large dispersion. The [N~II] temperature shows
           a substantially larger dispersion, but with all deviations being 
           within 1000~K; the majority of the objects give higher temperatures,
           causing a substantial offset. Electron densities are recovered 
           by 0.1 dex, with seven exceptions. Plasma parameters can be obtained
           from the published spectra for 53 objects. Helium abundances 
           are within 0.05~dex; three objects have lower abundances and
           two give higher values. The oxygen abundances show a strong
           scatter up to 0.4~dex without significant offset. The deviations
           in the N/O ratio are as large as 0.5~dex, with a clear tendency
           for HOPPLA getting larger values.
     \item the data of AC83 and AK87 display a similar behaviour:
           the temperatures (composed from several diagnostic ratios) 
           are within about 1000~K of the [O~III] temperatures in the reanalysis, 
           with 3 outliers. Electron densities show a strong scatter as well as 
           a systematic offset when compared with our values from [S~II], because 
           the published values are based on averages from different diagnostic
           line ratios. The helium abundances are systematically lower (by 
           about 0.05~dex) than in the reanalysis, with a scatter of the
           same magnitude. This offset is due to our inclusion of the
           corrections of the He~I emissivities for collisional excitation 
           based on Berrington \& Kingston (1988). Oxygen abundances show 
           significant scatter but without any offset. The N/O ratio has strong 
           dispersion with little offset.
  \end{itemize}

  Often the reason for one object to show a strong deviation in the results can be 
  traced to some particular cause, such as a diagnostic line ratio being at or close 
  to the limit of its sensitivity, resulting in a propagation of this error to other 
  quantities. It would go beyond the aims of this paper to discuss these objects in 
  detail, and thus trace in even more detail the origins of the differences found.

  \begin{table*}\centering
     \caption[]{Results of the reanalysis of PN spectra from previous studies
                using the HOPPLA code. The averages and the dispersions are 
                given for the differences of the original values and our results.
                Default values used by HOPPLA in the absence of diagnostic
                lines or in the high density limit are not taken into account.  
                Note that the electron temperatures and densities given by AC83 and AK87 
                are the values adopted from combining different diagnostic line ratios.
                The number of objects in each sample refers to those
                with helium abundances. For KB94 we also show the results based on the
                optical lines only (3700 ... 8000~\AA )}
     \label{t:reanalysis}
     \begin{tabular}{l|rr|rr|rr|rr|rr}
     \noalign{\smallskip} \hline \noalign{\smallskip}
          & \multicolumn{2}{|c|}{AC83} & \multicolumn{2}{|c|}{AK87} 
              & \multicolumn{2}{|c|}{KB94} & \multicolumn{2}{|c|}{KB94 (opt.)} & \multicolumn{2}{|c}{PSM01}  \\
      No.objects & \multicolumn{2}{|c|}{18} & \multicolumn{2}{|c|}{42} 
              & \multicolumn{2}{|c|}{52}   & \multicolumn{2}{|c|}{52} & \multicolumn{2}{|c}{31}  \\
          & mean & disp.& mean & disp.&  mean & disp.&  mean & disp. &  mean & disp. \\
    \noalign{\smallskip} \hline
     \noalign{\smallskip}
      T([OIII]) &($-260$) & (370) & (70)  & (410) & $-120$   & 230 & $-180$   & 220   & 20    & 60     \\
      T([NII])  &   ---   &   --- &  ---  & ---   & $-480$  & 430   & $-580$  & 320   & 10   & 100    \\
      lg(n[SII])& (0.073) &(0.155)& ($-0.060$)& (0.198) & $-0.003$ & 0.109 & $-0.003$ & 0.109 & 0.005 & 0.022  \\
     \noalign{\smallskip} 
      He/H      & 0.039 & 0.024 & 0.037 & 0.033 & 0.010 & 0.028 & 0.004 & 0.025 & 0.011 & 0.024  \\ 
      N/H       & $-0.072$ & 0.154 & 0.055 & 0.185 & $-0.157$ &  0.210  & $-0.104$ & 0.160 & $-0.043$ &  0.072 \\
      O/H       & 0.004 & 0.212 &  0.033 & 0.145 &$-0.041$ & 0.134 &$-0.009$ & 0.115 &$-0.040$ & 0.065  \\
      Ne/H      & 0.047 & 0.123 & 0.016 & 0.145 & $-0.055$ & 0.146 & $-0.022$ & 0.132 &  $-0.047$ & 0.061 \\
      S/H       & 0.063 & 0.162 &  $-0.011$ & 0.223 & $-0.067$ & 0.184  & $-0.088$ & 0.163 & --- &  --- \\
      Ar/H      & 0.196 & 0.139 &  0.197 & 0.167 & $-0.116$  &  0.199& 0.119 & 0.198 &   --- &   --- \\
      N/O       &$-0.102$ & 0.212 & 0.014 & 0.184 &$-0.120$ & 0.187 &$-0.092$ & 0.144 & $-0.002$ & 0.015  \\
     \noalign{\smallskip} \hline
     \end{tabular}
  \end{table*}
%
%
%
%
  \section{Comparison of the observational data}
  \label{s:comparedata}

  For the objects that we have in common with other works, we compare the 
  results from the reanalysis with HOPPLA of their spectra with the present 
  work. This reflects the differences in the measured intensities.

\begin{figure}[!ht]
      \includegraphics[trim = 70 70 70 70, scale=0.35, angle=270]{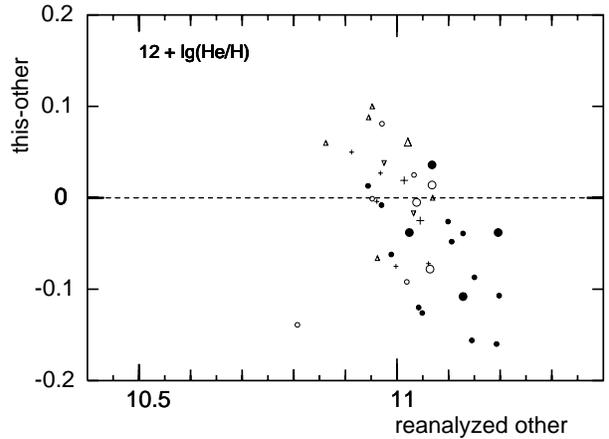} 
       \caption[]{Comparison of the helium abundances found in this work
                 with the values obtained by reanalysis of the data from the
                 other works. The symbols are the same as in Fig.\ref{f:cpaheii};
                 smaller symbols denote those objects which HOPPLA assumed
                 default values for the electron temperatures or density}
     \label{f:cpahe}   
   \end{figure}

  In Fig. \ref{f:cpahe} we compare our He/H abundances with the other studies.
  One notes a rather clear trend that we get lower helium abundances from our 
  own data than from the the data of PSM01, especially for higher abundances.
  The results from the other data also are in line with this trend, although
  the fewer numbers do not allow to make a firm conclusion.
  As shown in Fig. \ref{f:cpahei}, such a trend is already present in the 
  intensities of the He~I 5876~\AA\ line.

  However, if one compares the reanalyzed data of PSM01 with the other works,
  another clear correlation is found, as depicted in Fig. \ref{f:cpahe}.
  Again, this is reflected in PSM01 getting higher He~I intensities for
  the stronger lines than by previous studies (Fig. \ref{f:cppahei}). 

   \begin{figure}[!ht]
      \includegraphics[trim = 70 70 70 70, scale=0.35, angle=270]{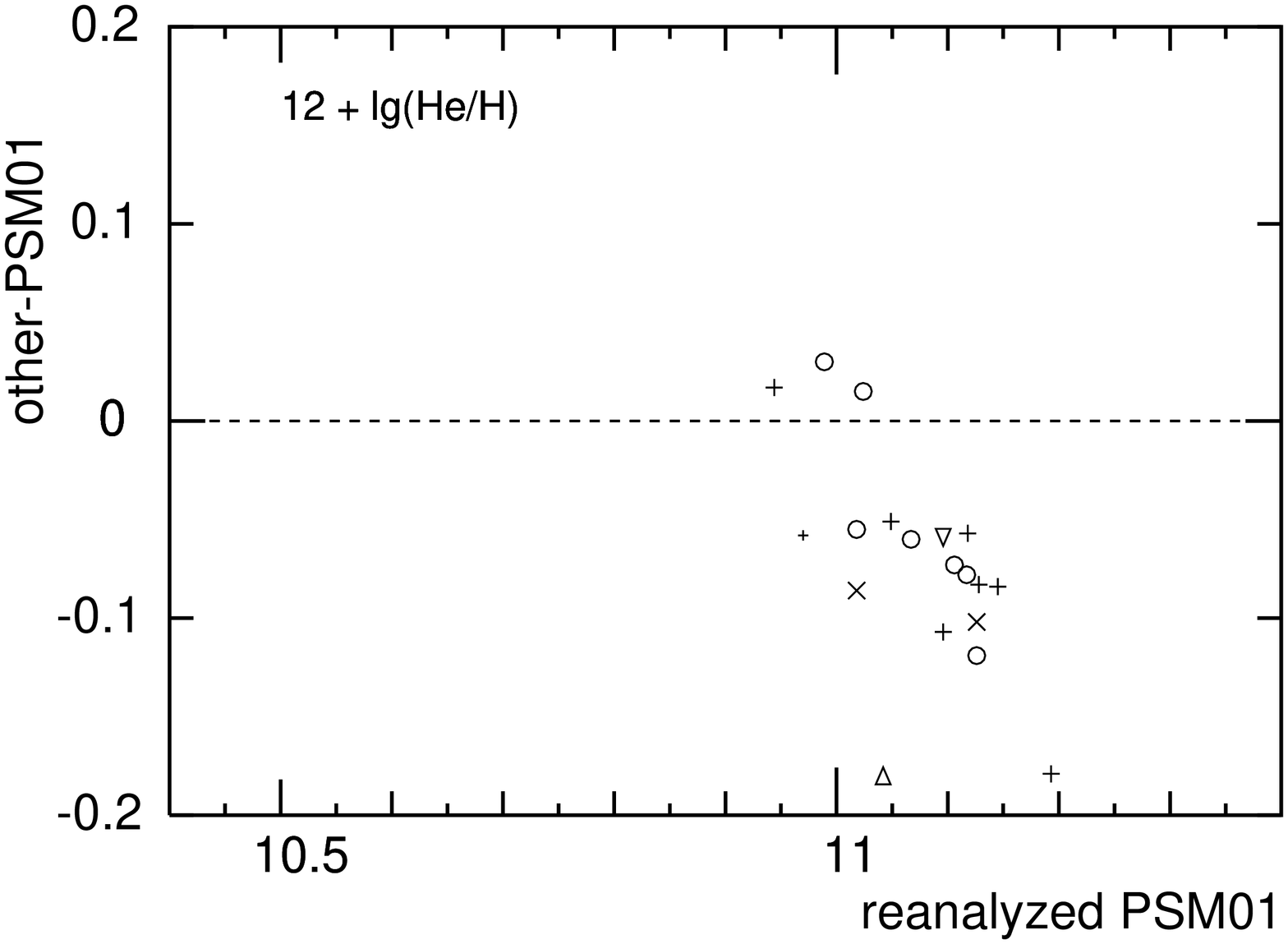} 
       \caption[]{Similar to Fig. \ref{f:cpahe}, but comparing the
                  reanalyzed PSM01 data with the reanalyzed data 
                  of AC83 ($\times$), AK87, KB94, CAK96, and CMAKS00, using the 
                  same symbols as in Fig.\ref{f:cpaheii}}
     \label{f:cppahe}   
   \end{figure}

   \begin{figure}[!ht]
      \includegraphics[trim = 70 70 70 70, scale=0.35, angle=270]{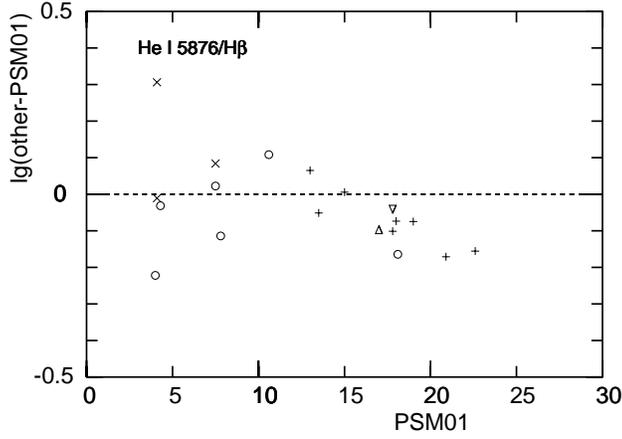} 
       \caption[]{Similar to Fig. \ref{f:cpahei}, but comparing the
                  PSM01 data with the data 
                  of AC83, AK87, KB94, CAK96, and CMAKS00, using the 
                  same symbols  as in Fig.\ref{f:cppahe}}
     \label{f:cppahei}   
   \end{figure}

  With the other elements, no such clear trend is found, partially
  because the scatter is substantially larger. We summarize the results
  as average and dispersion of the deviations found: Table \ref{t:thispena} 
  gives the comparison between our data and the reanalysis of the PSM01 
  data. The plasma parameters do not show a systematic offset, except for 
  the [N~II] temperatures, but a large scatter, which indicates
  differences in S/N ratio which affect the faint lines.
  The argon abundances obtained from the spectra 
   of PSM01 are substantially lower than from our spectra, because
   in PSM01 the intensity for the [Ar~III] 7135~\AA\ line - the dominant stage
   of ionization - are given only for three objects.
   
   In the abundances of N/H, O/H, Ne/H, and S/H there exist also offsets that
   are marginally acceptable, and one notes rather appreciable scatter.
   As the lower offsets and dispersions in the Ne/O and S/O ratios 
   indicate, the scatter originates from the O/H determinations and is translated 
   to the other elements by applying the ICFs. The scatter in O/H comes from 
   the electron temperature differences, and principally from differences in the 
   intensity of the [O~III] 4363~\AA\ line. As already seen in Fig.\ref{f:cpateo}, the 
   scatter in the [O~III] electron temperatures is much larger than the differences 
   due to the analysis methods. If we compare the reanalysed PSM01 data with 
   earlier data of the same objects, we find similar results; however, the 
   discrepances in the line intensities are stronger, as can be seen from
   Fig. \ref{f:cppahei}.

\begin{figure}[!ht]
     \includegraphics[trim = 70 70 70 70, scale=0.35, angle=270]{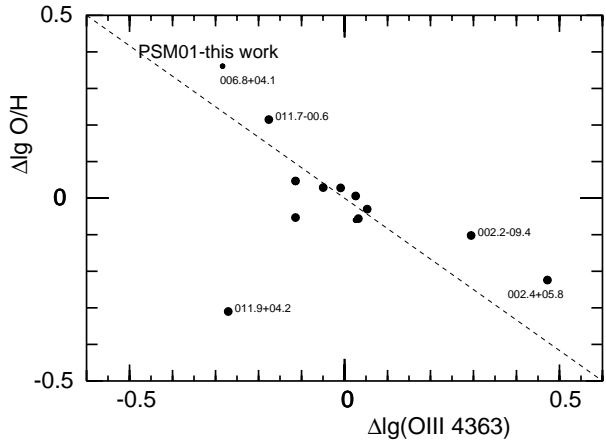} 
       \caption[]{The correlation between the deviations of oxygen abundances 
                  and intensities of the [O~III] 4363~\AA\ line for the nebulae 
                  in common with PSM01. The diagonal line indicates a linear
                  relation between the two deviations.}
     \label{f:epoiiioiii}   
  \end{figure}

   Let us consider this more closely in  Fig. \ref{f:epoiiioiii}: we compare the 
   differences between our values and PSM01 for the [O~III] 4363~\AA\ intensities 
   and the derived oxygen abundances. While for the majority of objects these 
   differences are less than 0.1 dex, outliers like PNG $006.8+04.1$ and 
   PNG $002.2+09.4$ exhibit [O~III] intensities that differ by as much as a 
   factor of 2, and their abundances differ correspondingly, as indicated by the
   diagonal line. PNG $006.8+04.1$ is one of the fainter objects of the sample
   which may thus be subject to a lower signal-to-noise rato. PNG $002.2+09.4$
   is one of the brighter ones, in whose spectrum no fault or anomality was
   noticed.

   The deviation of PNG $011.9+04.2$ has yet a different origin: the density 
   sensitive [S~II] line ratio is close to its high density limit. Thus the
   slightly lower intensity ratio measured in our spectra of 
   I(6717)/I(6731) = 0.50 (PSM01: 0.53) gives a significantly higher
   density of  9245 cm$^{-3}$ (2562). Due to the effects of collisional 
   deexcitation the nearly identical 5755~\AA\ line intensities 6.58 (7.10) 
   give a significantly [N~II] temperature 8576 K (9670 K), and hence
   higher O$^+$/H$^+$ ionic abundance $3.167\, 10^{-4}$ ($1.030\,10^{-4}$).
   This is more important than the change in [O~III] temperature
   10907 K (9965 K) from the larger [O~III] 4363~\AA\ intensity of  4.1 (2.2)!

  \begin{table}\centering
     \caption[]{The means and dispersions of the differences between the
                reanalyzed PSM01 data and this work, along with the formal
                error bars. For the abundances we use only values for which
                HOPPLA does not mark as uncertain, such as indicated by colons 
                in Table 3 (on-line version). For the line intensities, the differences of 
                $\lg(I/I({\rm H}\beta ))$ are used. The last column gives
                the number of objects.}
     \label{t:thispena}
     \begin{tabular}{lrrr}
     \noalign{\smallskip} \hline \noalign{\smallskip}
                   & mean & disp.& No.\\ 
    \noalign{\smallskip} \hline
     \noalign{\smallskip}
      T([OIII])     & $   -10  \pm 290   $  & $ 1130  \pm 210   $  & 15 \\
      T([NII])      & $   840  \pm 330   $  & $ 1100  \pm 240   $  & 11 \\
      lg(n[SII])    & $ -0.142 \pm 0.109 $  & $ 0.464 \pm 0.077 $  & 18 \\
     \noalign{\smallskip} 
      lg(He/H)      & $  0.053 \pm 0.019 $  & $ 0.078 \pm 0.013 $  & 17 \\
      lg(N/H)       & $  0.121 \pm  0.065 $  & $  0.242  \pm 0.046 $  & 14 \\
      lg(O/H)       & $ -0.001 \pm 0.043 $  & $ 0.161 \pm 0.031 $  & 14 \\
      lg(Ne/H)      & $-0.123 \pm 0.059  $  & $ 0.212  \pm 0.042 $  & 13 \\
      lg(S/H)       & $0.112 \pm 0.078   $  & $ 0.270 \pm 0.055 $  & 12 \\
      lg(Ar/H)      & $0.128 \pm 0.068$  & $0.096 \pm 0.048$  & 2\\
     \noalign{\smallskip} 
      lg(N/O)       & $ 0.122 \pm 0.054 $  & $0.203 \pm 0.038 $  & 14 \\
      lg(Ne/O)      & $ 0.055 \pm 0.020 $  & $0.073 \pm 0.014 $  & 13 \\
      lg(S/O)       & $-0.025 \pm 0.020 $  & $0.070 \pm 0.014 $  & 12 \\
     \noalign{\smallskip} 
  $\rm  [O\, III]\,5007$ & $  0.026 \pm 0.019 $  & $ 0.080 \pm 0.014 $  & 17 \\
  $\rm  He\, II\, 4686 $ & $ -0.091 \pm 0.082 $  & $ 0.218 \pm 0.058 $  &  7 \\
  $\rm  He\, I\, 5876 $  & $  0.056 \pm 0.015 $  & $ 0.064 \pm 0.011 $  & 17 \\
  $\rm [O\, III]\, 4363$ & $ -0.009 \pm 0.056 $  & $ 0.201 \pm 0.039 $  & 13 \\
  $\rm [S\, II]\, 6725$  & $  0.047 \pm 0.037 $  & $ 0.154 \pm 0.026 $  & 17 \\
     \noalign{\smallskip} \hline
     \end{tabular}
  \end{table}

\begin{figure}[!ht]
     \includegraphics[trim = 70 70 70 70, scale=0.35, angle=270]{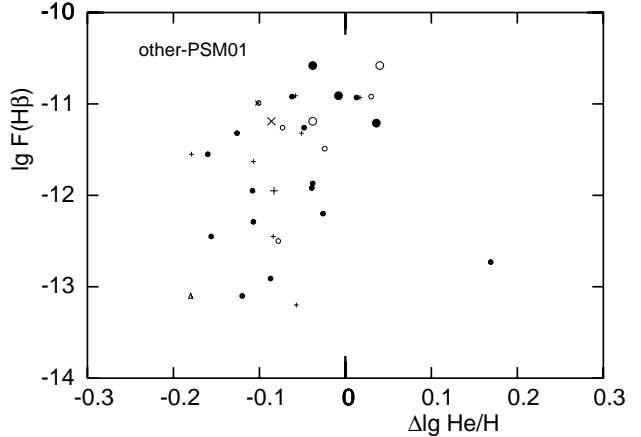} 
       \caption[]{Relation of the difference between helium
             abundances from the reanalyses of the data 
             of AC83 ($\times$), AK87($+$),  KB94($\circ$), 
             CAKS96 ($\bigtriangleup$), and this work ($\bullet$) 
             with those from PSM01. Smaller symbols indicate 
             nebulae where default values for electron temperature
             or density had to be assumed by HOPPLA.}
     \label{f:ehehflux}   
  \end{figure}

  In the quest to identify the origin of the deviation of the helium
  abundances derived by us and PSM01, the only clear trend is found to
  exist between this deviation and the absolute H$\beta$ flux, as presented
  in Fig. \ref{f:ehehflux}. We note that the results from the other works
  are in line with the results of this work, in that large negative 
  differences are found in faint objects. Because of the larger number of 
  objects in common, this trend is more evident with our data.

\begin{figure}[!ht]
      \includegraphics[trim = 70 70 70 70, scale=0.35, angle=270]{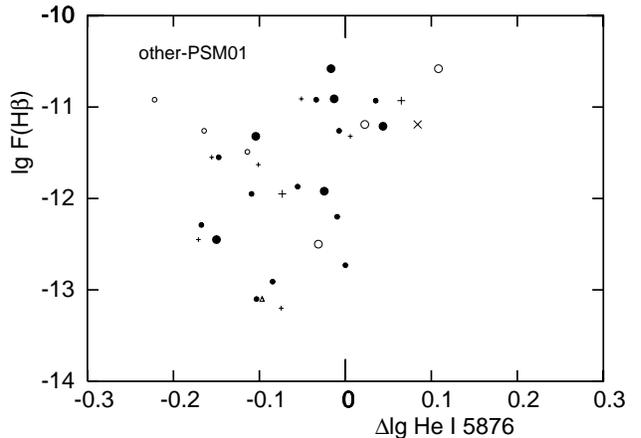} 
       \caption[]{Similar to Fig. \ref{f:ehehflux}, but for the
               intensity of the He~I 5876~\AA\ line.}
     \label{f:eheiflux}   
  \end{figure}

  Comparing the differences in the intensities of the He~I 5876\AA\ line 
  from the corresponding works (Fig \ref{f:eheiflux}), one notes that for 
  fainter nebulae the intensities from PSM01 tend to be smaller than in 
  the other works, including our own. Inspection of the [S~II] 6725~\AA\
  and [O~III] 4363~\AA\ lines, which are of comparable intensity, does
  not reveal a trend as clear as seen in the He~I line.

  We conclude that there exists a small but systematic difference in 
  the helium abundances between our work and PSM01, in that the
  nebular He~I lines in fainter objects have been either underestimated
  by us (and the other works) or overestimated by PSM01. Since the
  observational conditions are quite similar, yielding spectra of the
  same signal-to-noise ratio, we cannot determine the reason with 
  certainty. Because a thorough reanalysis of our data did not reveal
  a possible underestimation of He~I 5876~\AA\ fluxes, and the values
  are found to be in good agreement with the other He~I lines, we 
  tend to prefer our results.

  This difference is larger than the differences due to the analysis methods 
  and larger than uncertainties in atomic data, and thus it is somewhat 
  annoying; however, it is not large enough to affect the findings
  on the status of WRPN.
 
 \onecolumn
\begin{table}{Table 3. Elemental abundances for the nebulae. 
Uncertain values are marked with a colon. Q and EC are the same as in the Table 2.}
   \begin{tabular}{lrlllllllllll}
   \noalign{\smallskip} \hline \noalign{\smallskip}
    PN G        &  common name  & Spec.Type & Q & EC & \ \ He & \ \ \ N& \ \ \ O & \ \ Ne & \ \ \ S & \ \ Ar & \ \ Cl\\
   \noalign{\smallskip} \hline
   \noalign{\smallskip}
$000.4-01.9$& M 2-20	   &  WC5-6 & A &    4 &11.06~: & ~~8.17~~& ~~8.78~~& ~~7.85~~& ~~7.23~~& ~~6.61~:& ~~5.21~: \\
$002.2-09.4$& Cn 1-5	   &  WO4pe & A &    5 &11.10~~ & ~~8.52~~& ~~8.76~~& ~~8.24~~& ~~7.22~~& ~~6.68~~& ~~5.27~: \\
$002.4+05.8$& NGC 6369     &  WO3   & A &    5 &10.92~: & ~~8.29~~& ~~8.91~~& ~~8.40~~& ~~7.04~~& ~~6.37~:&  ---     \\
$003.1+02.9$& Hb 4	   &  WO3   & A &    6 &11.02~~ & ~~8.60~~& ~~8.72~~& ~~8.16~~& ~~7.08~~& ~~6.49~~& ~~5.17~: \\
$004.8-22.7$& He 2-436     &  WC4   & C &    5 &11.07~: & ~~7.29~:& ~~8.50~:& ~~7.75~:& ~~6.74~:& ~~5.86~:& ~~4.28~: \\
$004.9+04.9$& M 1-25	   &  WC4   & A &    4 &11.09~: & ~~8.41~~& ~~8.75~~& ~~7.47~~& ~~7.26~~& ~~6.66~~& ~~5.31~: \\
$006.0-03.6$& M 2-31	   &  WC4   & A &    5 &11.02~~ & ~~8.49~~& ~~8.70~~& ~~8.06~~& ~~7.04~~& ~~6.40~~& ~~4.97~: \\
$006.4+02.0$& M 1-31	   &  wels  & A &    5 &11.09~: & ~~8.31~~& ~~8.49~~&  ---    & ~~6.87~~& ~~6.42~:& ~~5.05~~ \\
   \noalign{\smallskip}
$006.8+04.1$& M 3-15	   &  WC4   & A &    5 &10.99~: & ~~8.01~~& ~~8.36~~& ~~7.68~~& ~~6.72~~& ~~6.14~:& ~~4.77~: \\
$009.4-05.0$& NGC 6629     &  wels  & A &    5 &10.96~~ & ~~7.56~~& ~~8.67~~& ~~7.96~~& ~~6.45~~& ~~6.23~:& ~~5.84~~ \\
$010.8-01.8$& NGC 6578     &  wels  & A &    5 &11.03~~ & ~~7.94~~& ~~8.75~~& ~~8.17~~& ~~6.88~~& ~~6.83~~& ~~5.20~~ \\
$011.7-00.6$& NGC 6567     &  wels  & A &    5 &10.96~~ & ~~7.61~~& ~~8.42~~& ~~7.67~~& ~~6.41~~& ~~5.73~~& ~~4.70~~ \\
$011.9+04.2$& M 1-32	   &  WO4pe & A &    4 &11.07~: & ~~8.35~~& ~~8.66~~& ~~7.58~~& ~~7.34~~& ~~6.89~~& ~~4.93~: \\
$012.2+04.9$& PM 1-188     &  WC10  & C &$<$ 2 & ---	& ~~8.12~:& ~~8.57~:&  ---    & ~~6.99~:& ~~5.84~:&  ---     \\
$016.4-01.9$& M 1-46	   &  wels  & A &$<$ 2 &10.88~~ & ~~7.95~~& ~~8.87~~&  ---    & ~~7.13~~& ~~6.53~:& ~~5.24~: \\
$017.9-04.8$& M 3-30	   &  WO1   & C &    7 &11.09~: & ~~7.02~:& ~~8.48~:&  ---    & ~~6.91~:& ~~6.53~:&  ---     \\
   \noalign{\smallskip}
$019.4-05.3$& M 1-61	   &  wels  & A &    5 &10.90~~ & ~~8.06~~& ~~8.49~~& ~~7.87~~& ~~6.58~~& ~~6.17~~& ~~5.48~: \\
$019.7-04.5$& M 1-60	   &  WC4   & A &    5 &11.07~~ & ~~8.96~~& ~~8.78~~& ~~8.22~~& ~~7.21~~& ~~6.67~~& ~~4.72~: \\
$020.9-01.1$& M 1-51	   &  WO4pe & A &    5 &11.08~: & ~~8.29~~& ~~8.92~~& ~~8.52~~& ~~7.28~~& ~~6.54~:& ~~5.24~: \\
$027.6+04.2$& M 2-43	   &  WC7-8 & C &    4 &10.92~: & ~~7.24~:& ~~8.63~:& ~~7.10~:& ~~6.87~:& ~~6.16~:& ~~4.55~: \\
$029.2-05.9$& NGC 6751     &  WO4   & A &    5 &11.06~: & ~~8.34~~& ~~8.69~~& ~~8.06~~& ~~6.94~~& ~~6.46~:&  ---     \\
$034.6+11.8$& NGC 6572     &  wels  & A &    5 &11.04~~ & ~~8.31~~& ~~8.60~~& ~~7.93~~& ~~6.52~~& ~~6.33~~& ~~4.83~~ \\
$038.2+12.0$& Cn 3-1	   &  wels  & A &$<$ 2 &10.68~~ & ~~7.95~~& ~~8.79~~&  ---    & ~~6.99~~& ~~6.15~~& ~~4.95~: \\
$048.7+01.9$& He 2-429     &  WC4   & A &    4 &11.06~: & ~~8.41~~& ~~8.77~~&  ---    & ~~7.24~~& ~~6.81~~& ~~5.34~: \\
   \noalign{\smallskip}
$055.5-00.5$& M 1-71	   &  wels  & A &    5 &11.03~: & ~~8.62~~& ~~8.76~~& ~~8.01~~& ~~6.52~~& ~~6.50~~& ~~4.93~~ \\
$057.2-08.9$& NGC 6879     &  wels  & A &    5 &10.99~~ & ~~7.93~~& ~~8.52~~& ~~7.87~~& ~~6.80~~& ~~6.14~~& ~~4.64~: \\
$061.4-09.5$& NGC 6905     &  WO2   & B &    7 &10.93~: & ~~8.47~~& ~~8.85~~& ~~8.18~~& ~~7.13~~& ~~6.18~~& ~~5.60~: \\
$068.3-02.7$& He 2-459     &  WC9   & C &$<$ 2 &~~9.48~:& ~~7.77~:& ~~7.85~:&  ---    & ~~7.03~:&  ---    & ~~4.45~: \\
$253.9+05.7$& M 3-6	   &  wels  & A &    5 &11.05~~ & ~~7.77~~& ~~8.71~~& ~~8.15~~& ~~6.73~~& ~~6.35~~& ~~5.07~: \\
$258.1-00.3$& He 2-9	   &  wels  & A &    4 &10.96~: & ~~7.73~~& ~~8.44~~& ~~7.69~~& ~~6.57~~& ~~6.80~~& ~~4.70~: \\
$274.6+02.1$& He 2-35	   &  wels  & A &    5 &11.00~~ & ~~7.23~~& ~~8.73~~& ~~8.10~~& ~~6.77~~& ~~6.24~~& ~~4.95~: \\
$278.1-05.9$& NGC 2867     &  WO2   & A &    7 &10.99~~ & ~~7.99~~& ~~8.64~~& ~~7.96~~& ~~6.71~~& ~~6.04~~& ~~4.97~: \\
   \noalign{\smallskip}
$278.8+04.9$& PB 6	   &  WO1   & A &    7 &11.16~~ & ~~8.94~~& ~~8.79~~& ~~8.23~~& ~~6.82~~& ~~6.12~~&  ---     \\
$285.4+01.5$& Pe 1-1	   &  WO4   & A &    5 &10.99~: & ~~8.15~~& ~~8.62~~& ~~8.02~~& ~~6.62~~& ~~6.34~~& ~~4.91~~ \\
$291.3-26.2$& Vo 1	   &  WC10  & C &$<$ 2 &11.02~: & ~~7.39~:& ~~8.65~:&  ---    & ~~6.58~:&  ---    &  ---     \\
$292.4+04.1$& PB 8	   &  WC5-6 & A &    6 &11.08~~ & ~~8.33~~& ~~8.82~~& ~~8.18~~& ~~6.84~~& ~~6.69~~& ~~5.36~~ \\
$300.7-02.0$& He 2-86	   &  WC4   & A &    5 &11.05~: & ~~8.79~~& ~~8.67~~& ~~7.93~~& ~~7.13~~& ~~6.54~~& ~~5.13~: \\
$307.2-03.4$& NGC 5189     &  WO1   & A &    5 &10.72~: & ~~8.41~~& ~~8.38~~& ~~7.75~~& ~~7.53~~& ~~6.41~~& ~~4.85~: \\
$327.1-02.2$& He 2-142     &  WC9   & B &$<$ 2 &10.67~: & ~~8.22~:& ~~8.95~:&  ---    & ~~6.90~:& ~~5.40~:& ~~5.12~: \\
$331.3+16.8$& NGC 5873     &  wels  & A &    7 &10.95~~ & ~~7.22~~& ~~8.38~~& ~~7.74~~& ~~7.23~:& ~~5.67~:& ~~5.18~~ \\
   \noalign{\smallskip}
$336.2-06.9$& PC 14	   &  WO4   & A &    5 &11.03~~ & ~~8.25~~& ~~8.77~~& ~~8.12~~& ~~7.06~~& ~~6.38~~& ~~5.09~: \\
$337.4+01.6$& Pe 1-7	   &  WC9   & B &$<$ 2 &10.17~: & ~~7.22~:& ~~6.93~:&  ---    & ~~6.10~:& ~~4.97~:& ~~3.82~: \\
$351.1+04.8$& M 1-19	   &  wels  & A &    4 &11.01~: & ~~8.39~~& ~~8.86~~& ~~8.17~~& ~~7.06~~& ~~6.62~:& ~~5.28~: \\
$355.2-02.5$& H 1-29	   &  WC4   & A &    5 &11.01~: & ~~8.26~~& ~~8.67~~& ~~8.04~~& ~~6.81~~& ~~6.29~~& ~~4.95~: \\
$355.9-04.2$& M 1-30	   &  wels  & A &    3 &11.09~~ & ~~8.49~~& ~~8.79~~& ~~7.89~~& ~~7.30~~& ~~6.76~~& ~~5.28~: \\
$356.7-04.8$& H 1-41	   &  wels  & A &    7 &11.03~~ & ~~8.00~~& ~~8.57~~& ~~7.87~~& ~~6.78~~& ~~6.09~~& ~~5.27~~ \\
$357.1+03.6$& M 3-7	   &  wels  & A &    4 &11.00~~ & ~~7.91~~& ~~8.68~~& ~~7.98~~& ~~7.06~~& ~~6.36~:& ~~5.11~: \\
$358.3-21.6$& IC 1297	   &  WO3   & A &    7 &11.05~~ & ~~8.37~~& ~~8.77~~& ~~8.10~~& ~~7.10~~& ~~6.27~~& ~~5.37~~ \\
   \noalign{\smallskip} \hline
   \end{tabular}
\end{table}

\end{document}